\documentclass[12pt]{iopart}

\expandafter\let\csname equation*\endcsname\relax

\expandafter\let\csname endequation*\endcsname\relax
\usepackage{iopams}
\usepackage{amsmath} 
\usepackage{graphicx}
\usepackage{dcolumn}
\usepackage{bm}
\usepackage{epstopdf}
\usepackage{array}
\usepackage{braket}
\usepackage{xcolor}
\usepackage{tcolorbox}
\usepackage[breaklinks=true,colorlinks]{hyperref}
\begin{document}

\title{Thermodynamic sensing of quantum nonlinear noise correlations}

\author{Nilakantha Meher}
\address{AMOS and Department of Chemical and Biological Physics,
Weizmann Institute of Science, Rehovot 7610001, Israel}
\address{Department of Physics, SRM University-AP, Amaravati, Andhra Pradesh 522240, India}
\ead{nilakantha.meher6@gmail.com}
\author{Tom\'{a}\v{s} Opatrn\'y}
\address{Department of Optics, Faculty of Science, Palack\'y University, 17. listopadu 50, 77146 Olomouc, Czech Republic}
\author{Gershon Kurizki}
\address{AMOS and Department of Chemical and Biological Physics,
Weizmann Institute of Science, Rehovot 7610001, Israel}

\begin{abstract}
We put forth the concept of quantum noise sensing in nonlinear two-mode interferometers coupled to mechanical oscillators.  These autonomous machines  are capable of sensing quantum nonlinear correlations of two-mode noisy fields  via their thermodynamic variable of extractable work, alias work capacity or ergotropy. The fields are formed by thermal noise input via its interaction with multi-level systems inside the interferometer. Such interactions amount to the generation of two-mode \textit{quantum nonlinear gauge fields} that may be partly unknown. We show that by monitoring a mechanical oscillator coupled to the interferometer, one can sense the work capacity of one of the output field modes and thereby reveal the quantum nonlinear correlations of the field. The proposed quantum sensing method can provide an alternative to quantum multiport interferometry where the output field is unraveled by tomography. This method may advance the simulation and control of multimode quantum nonlinear gauge fields. 
\end{abstract}

%
%
%
%
%

\section{Introduction}
In this paper we establish a connection between two hitherto unrelated areas: \textit{quantum sensing of noisy nonlinear fields} and their \textit{thermodynamic variables}. 
The goal of quantum sensing is to maximize signals that carry information on quantum systems or processes while suppressing the background noise \cite{Degen2017RevModPhys,Kurizki2015PNAS,kurizki2022Book}. This goal, however, ignores the unique \textit{information contained in the noise:} paradoxically, in certain situations \textquotedblleft\textit{the noise is the signal}\textquotedblright~ \cite{Landauer1998Nature}. In particular, noise correlations are at the heart of spin  spectroscopy \cite{Henry1996RMP,Zapasskii2013AdvOptPhoton,Zwick2016PRAppl,Alvarez2011PRL,Smith2012PNAS,
Staudenmaier2011Arx}. In emerging quantum technologies \cite{Kurizki2015PNAS}, the ability to  sense noisy fields of light  
with \textit{apriori unknown } statistical properties \cite{VirziPRL2022,HANBURYBROWN1956,Scully1988PRL,Short1983PRL,Carmichael_BOOK,Gardiner,
ScullyZubairy,Garraway1994PRA,Lounis2005RepProgPhys, Buckley2012RepProgPhys,Meher2018QInP,Meher2022EPJP, Goldman2014PRX, delPino2023PRL}  are central priorities.  

The existing arsenal of sensing noisy quantum fields consists of recording correlations of two or more detectors at the output of interferometers and analyzing this output tomographically \cite{Chuang1997JModOpt, Arino2003AdvImagElecPhys,Mohseni2008PRA}. The simplest task is to distinguish thermal (Gaussian) from superthermal or sub-Poissonian (non-classical) fluctuations \cite{HANBURYBROWN1956,Scully1988PRL,Short1983PRL,Carmichael_BOOK,Gardiner,
ScullyZubairy,Garraway1994PRA,Lounis2005RepProgPhys, Buckley2012RepProgPhys,Meher2018QInP,Meher2022EPJP, Hong1987PRL, Kwiat1995PRL,Bai2017JOSAB, Imamoglu1997PRL,Auffeves2011NJP, Kazimierczuk2015PRL,Albert2011NatComm,Leymann2015PRApplied,
Redlich2016NJP,
Auffeves2011NJP,Tureci2006PRA,Christos2018PNAS, Kazimierczuk2015PRL,Grujic2013PRA}. Yet,  particularly for multi-mode fields, the complexity of the underlying quantum nonlinear processes \cite{Imamoglu1997PRL,Auffeves2011NJP, Kazimierczuk2015PRL,Albert2011NatComm,Leymann2015PRApplied,
Redlich2016NJP,
Auffeves2011NJP,Tureci2006PRA,Christos2018PNAS, Kazimierczuk2015PRL,Grujic2013PRA} often makes their detailed interferometric noise signature hard to record and interpret. Since second-order correlations in general do not fully identify a multi-mode nonlinear process, \textit{high-order noise correlations} need to be measured, such as those predicted by the Scully-Lamb quantum theory of the laser \cite{Peng2021FP}. To this end, the multi-port photon statistics of the interferometric output must first be recorded by high-efficiency photon-number sensitive detectors. {Then, laborious quantum process tomography must be performed by measuring the output correlations for a large set of input states, thereby reconstructing the underlying process \cite{Chuang1997JModOpt, Arino2003AdvImagElecPhys,Lvovsky2009RMP}. Full tomography of such quantum processes requires a vast quantum resource overhead \cite{Chapman2022OptExp}}. 

Here we address a fundamental, hitherto unexplored, question: \textit{can the quantum statistics of a noisy multi-mode field be sensed via its thermodynamic variables?}  We show unequivocally that the thermodynamic variable of extractable work, alias work capacity (WC) or ergotropy \cite{kurizki2022Book,Allahverdyan_2000,PRE2014,Ghosh2018,
 niedenzu18quantum, Francica2017npjQI}, provides a clear signature of nonlinear two-mode quantum noise statistics formed in the interferometer. It can thus reliably sense quantum nonlinear processes arising inside the interferometer, which may be viewed as a \textquotedblleft black-box". { Among them, of high interest are complex processes that engender only partly controllable quantum electrodynamic gauge fields \cite{Goldman2014PRX,delPino2023PRL}. These fields must be unraveled to be simulated, posing a challenge to quantum sensing. } 

{In contrast to the existing methods, the sensing of quantum nonlinear correlations in the \textquotedblleft black-box" via the output WC requires monitoring a single-port output of an interferometer, rather than two-port or multi-port correlations. 
This monitoring is simpler than full quantum-process tomography, since it only seeks partial information on the probed process, just enough to distinguish it from other possible processes in the \textquotedblleft black-box".}

{Such two-mode quantum nonlinear processes can only generate gauge fields by (i) off-resonant photon exchange to some order $s$, resulting in phase correlation of the two modes; (ii) $k$-th order resonant photon exchange between the modes; or (iii) combinations thereof. If we venture beyond the foregoing two-mode autonomous unitary processes, we may similarly consider parametric processes \cite{Ghosh2017} or measurement induced nonlinear processes \cite{Opatrny2000PRA,OpatrnyPRL21}}  

\begin{figure}
\begin{center}
\includegraphics[height=6cm,width=7cm]{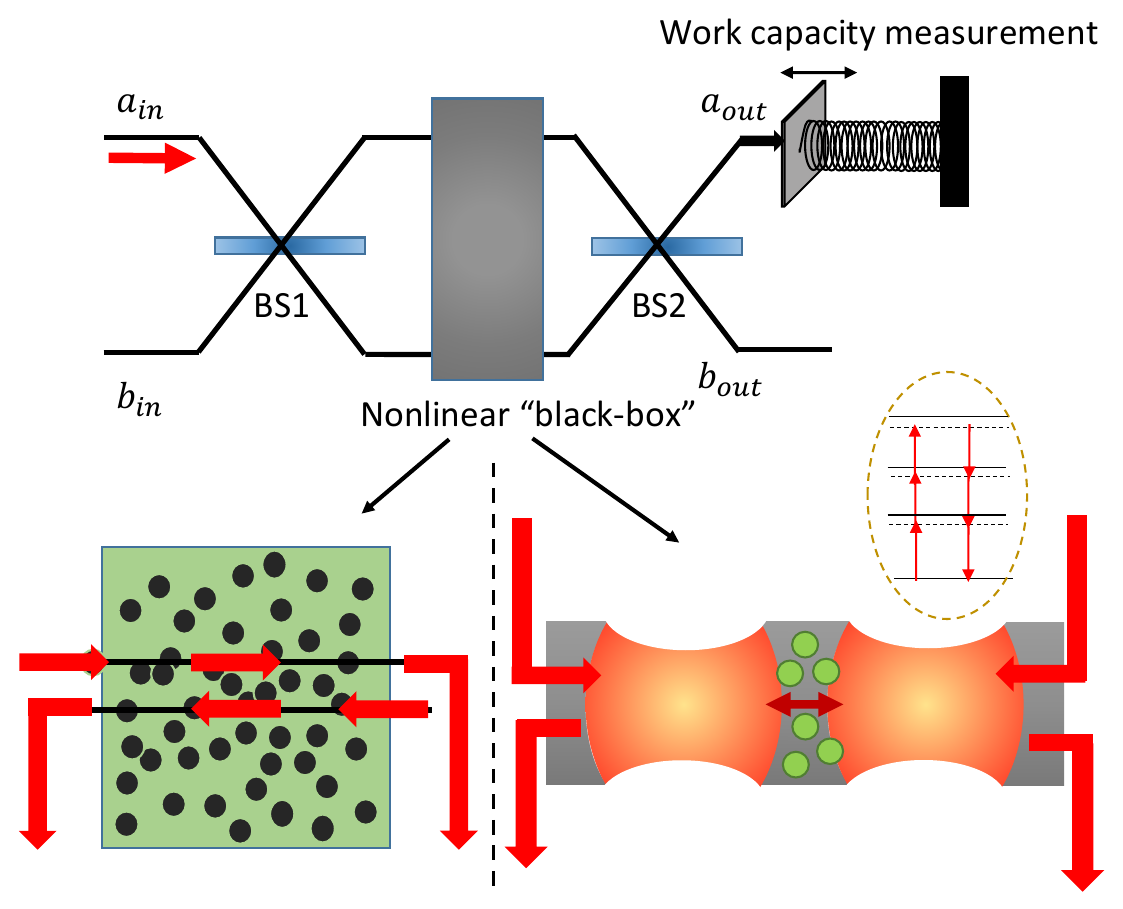}
\end{center}
\caption{{ A Mach-Zehnder interferometer that consists of two 50:50 beam splitters and a nonlinear \textquotedblleft black-box" coupler causing either cross-phase dispersive correlation or multi-photon exchange of two field modes. The interferometer is injected  with two independent thermal input fields and coupled to a mechanical oscillator that serves for work capacity measurement. A \textquotedblleft black-box"  with strong two-mode nonlinear coupling may contain cold-atom polariton collisions (left) or cavity-mode fields strongly coupled by atoms, molecules or artificial atoms (right). } }\label{NLMZI}
\end{figure}

Our analysis establishes the following general principles concerning the unique characterization of nonlinear two-mode processes by \textit{single-mode} WC: (1) If one input mode of the interferometer is empty and the other carries noise in a \textit{passive (typically, thermal) state}, void of WC \cite{kurizki2022Book,Allahverdyan_2000,PRE2014,Ghosh2018,
 niedenzu18quantum, Francica2017npjQI}, then only nonlinear interactions in the interferometer can filter this noise and render the individual single-mode output states non-passive/non-Gaussian, endowed with WC. (2) Different nonlinear interactions yield output states with qualitatively different time-dependence and efficiency of the WC, thus providing a distinct signature of the interaction. (3) This signature is particularly transparent for low photon-number input, wherein the single-mode output WC is reciprocal to the correlation (coherence) functions of the output. {(4) The WC that stores relevant information on the field state cannot be measured directly, not being a canonical variable \cite{kurizki2022Book}. Yet, it can be probed, e.g., by coupling it optomechanically to a mechanical oscillator \cite{Aspelmeyer2014RevModPhys, Treutlein2014NatNanotech,Chan2011Nature, Schliesser2009NatPhys, Bild2023Science} (Fig. \ref{NLMZI}), the procedure being akin to homodyning \cite{Lounis2005RepProgPhys,OpatrnyPRL21}. }

{We organize our article as follows: We introduce our interferometric setup in Sec. \ref{Scenario} and calculate the time-dependence and efficiency of the output WC for different nonlinear processes in Sec. \ref{Workcapacity}. In Sec. \ref{Coherencefunction}, we show that the interferometer can sense the quantum fluctuations of nolinearly correlated two-mode fields generated from different nonlinear processes. In Sec. \ref{optomechanics}, we infer the WC of the output field by coupling it to a mechanical oscillator. The results and the outlook are summarized in Sec. \ref{discussion}.}

\begin{figure}
\begin{center}
\includegraphics[height=5.5cm,width=7cm]{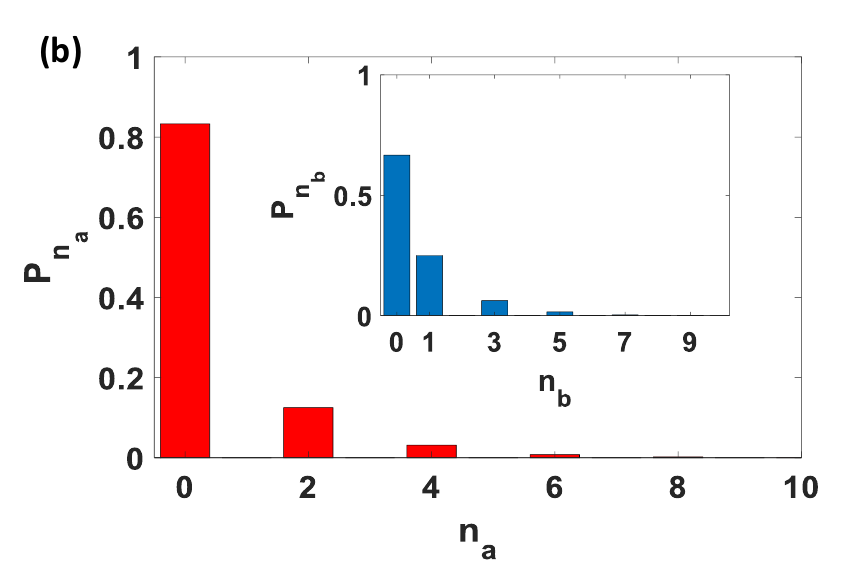}
\end{center}
\caption{{ Photon-number distribution in the output state $\rho_{a|out}$ of an interferometer containing a cross-Kerr coupler with $\chi t=\pi$.  Inset:  same for the output state $\rho_{b|out}$. The input to mode $a$ is thermal state with $\bar n_a=1$ and the mode $b$ is in vacuum with $\bar n_b=0$.} }\label{CKPaPb}
\end{figure}

\begin{figure*}
\begin{center}
\includegraphics[height=5.5cm,width=14cm]{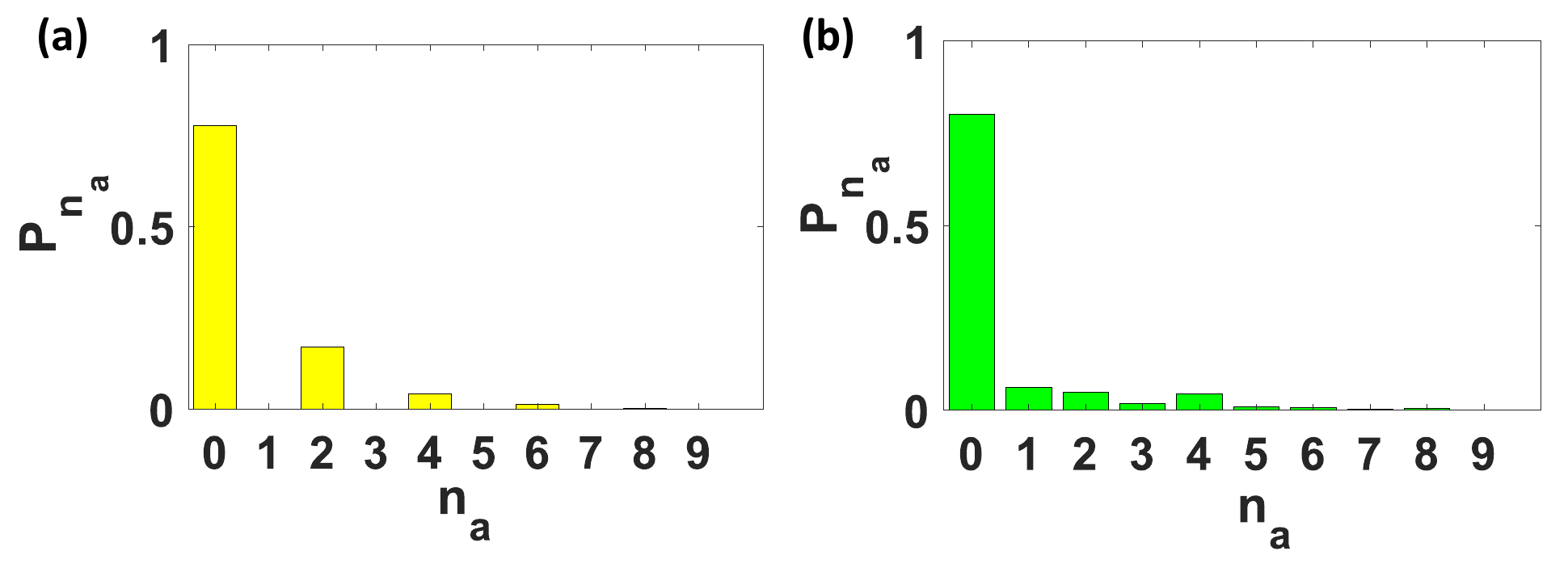}
\end{center}
\caption{{ (a) Photon number distribution in the state $\rho_{a|out}$ for an interferometer containing 2-photon exchange coupler with $gt=\pi$. (b) Same for 3-photon exchange coupler. The input to mode $a$ is thermal state with $\bar n_a=1$ and the mode $b$ is in vacuum with $\bar n_b=0$. } }\label{k23PaPb}
\end{figure*}

\section{Scenario}\label{Scenario}
We consider a simple two-arm Mach-Zehnder interferometer (MZI) wherein two noisy quantized input modes interfere on a 50/50 beam splitter (BS) and subsequently merge in an element (a cavity, a microsphere or a cell) that mixes them via  nonlinear interactions [Fig. \ref{NLMZI}]. The quantized nonlinear interaction may be complicated or even untractable \cite{Henry1996RMP,
Carmichael_BOOK,ScullyZubairy,Gardiner,Imamoglu1997PRL,Auffeves2011NJP, Kazimierczuk2015PRL,Albert2011NatComm,Leymann2015PRApplied,
Redlich2016NJP,Auffeves2011NJP, Kazimierczuk2015PRL,Grujic2013PRA}, and may be assumed \textit{unknown} to us, \textit{i.e.,} a \textquotedblleft \textit{black-box}" \cite{Chuang1997JModOpt}.

The overall two-mode transformation of the input state $\rho_{in}$ to the interferometer output state $\rho_{out}$ is 
\begin{align}
\rho_{out}=\mathcal{L}_D (\hat U\rho_{in}\hat U^\dagger),
\end{align}
Here, $\mathcal{L}_D$ is the superoperator that accounts for dissipation or decoherence in the medium and $\hat U=\hat U_{L1}\hat U_{NL}\hat U_{L2}$ is a unitary transformation where 
$\hat U_{L1(L2)}$ are linear sequential transformations (caused by BS1(2) in Fig. \ref{NLMZI}) and $\hat U_{NL}$ is a \textit{nonlinear transformation} in the 2-mode space. Under the assumption that the nonlinear MZI transformation occurs over time/length $(t \approx l/c)$ that is far shorter than its coherence time/length, $\mathcal{L}_D$ effects can be neglected. {The input noise is thus the dominant source of incoherence. }

 
We have recently shown \cite{Opatrny2023ScAdv} that the crux of the difference between linear and nonlinear transformations, $\hat U_{L}$ and $\hat U_{NL}$, in multimode interferometers, is that $\hat U_{L}$ cannot change the passivity (lack of WC) of the output modes. By contrast, a \textit{nonlinear} $\hat U_{NL}$ can render each output mode non-passive, i.e.,  endowed with WC. This difference is here shown to make \textit{thermal input distinctly advantageous as a sensing resource} compared to pure-state input: thermal noise (having zero WC/ergotropy) acquires WC only when transformed by $\hat U_{NL}$ and therefore can sensitively unravel the nonlinear process, whereas the WC of pure state input is typically non-zero (for Fock and coherent states) and thus is less drastically changed by $\hat U_{NL}$. 

We therefore assume a \textit{passive, e.g., thermal input state} in mode $a$, which means $
\rho_{a}=\sum_{n}P_n\ket{n}\bra{n}$ with monotonically decreasing probabilities $P_n\geq P_{n+1}$. For simplicity, we take mode $b$ to be empty, $\rho_{b}=\ket{0}\bra{0}$.

We here show that WC can serve as a signature of gauge fields generated in a \textquotedblleft black-box" consisting of field modes interacting with matter upon eliminating the matter degrees of freedom. All possible gauge-field Hamiltonians that can be autonomously generated (without external drive) are nonlinear functions of two-mode Stokes (pseudospin) operators 
\begin{align}
\hat J_x&=\frac{1}{2}(\hat a^\dagger \hat b+\hat a \hat b^\dagger),\\
\hat J_y&=\frac{-i}{2}(\hat a^\dagger \hat b-\hat a \hat b^\dagger),\\
\hat J_z&=\frac{1}{2}(\hat a^\dagger \hat a - \hat b^\dagger \hat b),
\end{align}
where $\hat a$ and $\hat b$ are the annihilation operators of two coupled frequency-degenerate modes (extension to non-degenerate modes does not provide new insights). These are energy conserving processes, and hence, $\hat{N}=\hat a^\dagger \hat a + \hat b^\dagger \hat b$ is an invariant quantity. The possible generic forms of such two-mode nonlinear interactions in terms of Stokes operators as follows:\\ 
\noindent 
\textbf{(1) Nonlinear dispersive (cross-phase) coupling}: In its simplest form, this coupling is a cross-Kerr Hamiltonian \cite{Carmichael_BOOK,Gardiner,
ScullyZubairy}  that correlates the degenerate two-mode phase shifts via 
\begin{align}
\hat U_{NL}(\chi t)=e^{-i\chi t \hat n_a \hat n_b},
\end{align} 
where $\hat n_{a(b)}$ are  the respective number operators of modes $a$ and $b$; {and the parameter $\chi$ is the cross-Kerr phase-shift when one photon in each mode passes  through the nonlinear medium in unit time.} The exponent  $\hat n_a \hat n_b$ can be expressed as 
\begin{align}
\hat n_a \hat n_b=\tfrac{1}{4}\hat N^2-\hat J_z^2.
\end{align}

{Higher $(s>1)$ order dispersive cross-phase couplings, represented by 
\begin{align}
\hat U_{NL}^{(s)}(\chi t)=\exp\left[{-i\chi t \hat n_a^s \hat n_b^s}\right],
\end{align}
involve higher powers of $\hat J_z$ and $\hat N$. All such processes transform Gaussian passive input to a non-passive non-Gaussian output \cite{kurizki2022Book, Allahverdyan_2000,PRE2014,Ghosh2018,
 niedenzu18quantum, Francica2017npjQI,Opatrny2023ScAdv}) in each port of the interferometer, as shown in Fig. \ref{CKPaPb}.} 

\textbf{(2) Multi-photon exchange process:}
Nonlinear multi-order or multi-wave mixing \cite{Zuo2006PRL, boyd}, wherein two frequency-degenerate modes exchange $k$ photons simultaneously, is governed by the unitary operator 
\begin{align}
\hat U_{NL}^{(k)}=e^{-igt(\hat a^{\dagger k} \hat b^k+\hat a^k \hat b^{\dagger k})},
\end{align}
$g$ being the coupling strength with which the two modes exchange photons. The exponent is expressible in terms of Stokes operators as: 
\begin{align}
\hat a^{\dagger k} \hat b^k+\hat a^k \hat b^{\dagger k}=(\hat J_x+i\hat J_y)^k+(\hat J_x-i\hat J_y)^k.
\end{align}  
For $k=1$, the exponent scales as $\hat J_x$ (linear beam splitter transformation or parametric Raman exchange \cite{Jouravlev2004PRA}) and the interferometer output remains passive, without merit for this study. By contrast, here we consider $k>1$ photon exchange processes that are non-Gaussian operations and change the photon number distribution to be non-monotonic, hence non-passive \cite{Allahverdyan_2000,PRE2014,Ghosh2018,
 niedenzu18quantum, Francica2017npjQI}. { Following the general approach of nonlinear optics \cite{boyd}, the response of the medium to electromagnetic fields is expanded in ascending powers of the field operators, and correspondingly consecutive orders of  photon-number exchange \cite{drummond2014quantum}. This  perturbative expansion can be  safely cut at the $k=4$ order of photon exchange (8th  order in the field operators) since higher-order processes are practically negligible in the quantum domain of small photon numbers.}


\noindent
\textbf{(3) Hybrid dispersive and exchange process:}  In principle, such processes can be generated by Hamiltonians containing cross products of $\hat J_x, \hat J_y$ and $\hat J_z$ to some power. These processes will not be considered here, but can be characterized by combining the methods described below for type (1) and (2) processes. {For hybrid processes, namely, processes that are characterized by mixed parity of photon exchange, it may be possible to combine the WC characteristics  for even- or  odd-order photon exchange discussed here,  provided parity selection is achieved by the dominant  process whereas the weaker process diminishes this parity selection.   }

\section{Work capacity as quantum fluctuation sensor}\label{Workcapacity}
We now show that the \textit{ time-dependence and efficiency} of the output WC can provide a specific signature of the nonlinear process:\\
\noindent
(a) {In the case of interferometers containing either \textbf{cross-Kerr or even-$k$ ($k=2,4)$ photon exchange couplers}, the interferometer transformation of the Hamiltonian generates two-photon exchange [SI Sec. 2]. The output state of mode-$a$ then acquires the general form 
\begin{align}\label{RhoAout}
\rho_{a|out}=\tilde{P}_0(t)\ket{0}\bra{0}+\sum_{n~even}\tilde{P}_n(t) \ket{n}\bra{n},
\end{align}
in which the \textit{odd number states $(n=1,3,...)$ completely cancel out by destructive interference} [Figs. \ref{CKPaPb} and \ref{k23PaPb}(a)]. This parity filtering renders the output photon-number distribution non-monotonic (non-passive) and non-Gaussian}. Notably, for cross-Kerr coupler with $\chi t=(2m+1)\pi$, $m$ being an integer, the two output modes $a_{out}$ and $b_{out}$ exhibit complementary trends, whereby, $a_{out}$ contains only even number of photons and $b_{out}$ contains only odd number of photons [inset of Fig. \ref{CKPaPb}]. 
  
The resulting output WC for {cross-Kerr or even-$k$ photon exchange couplers} can be shown to be [SI Sec. 2]
\begin{align}\label{ergotropy}
\langle W_{a|out}(t) \rangle=\frac{\sum_{n~even} n\tilde{P}_n(t)}{2}=\frac{\langle \hat a^{\dagger}_{out}\hat a_{out}\rangle}{2}.
\end{align}
Here the output of mode $a$ is described by the annihilation and creation operators $\hat a_{out}$ and $\hat a^\dagger_{out}$. The WC is thus half of the output average photon number in mode $a_{out}$. {The other half does not produce WC but only entropy and hence, amounts to heat production.} 

\begin{figure}[t]
\begin{center}
\includegraphics[height=6cm,width=7.5cm]{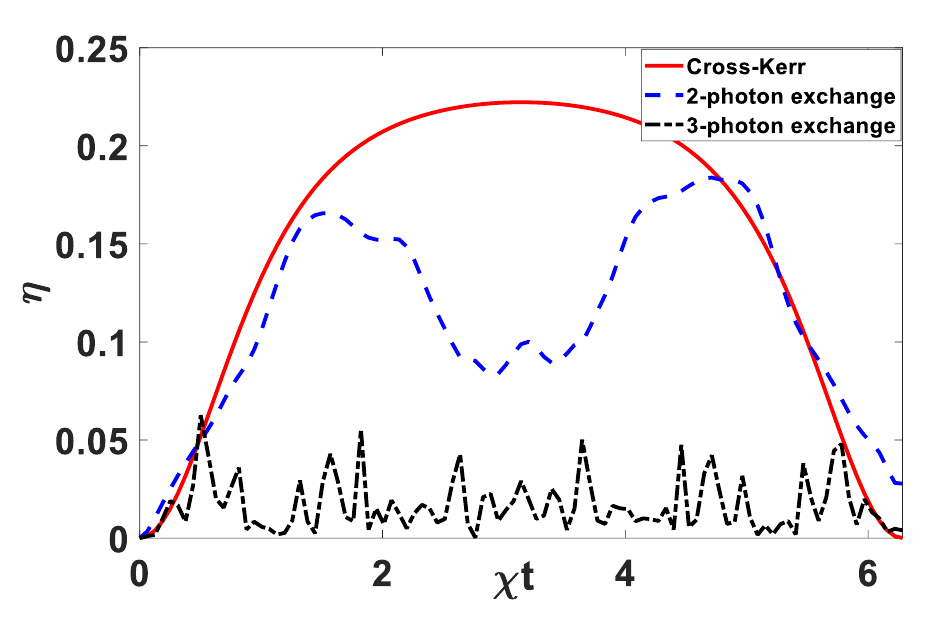}
\end{center}
\caption{{ Efficiency of output WC normalized by the input average photon number, $\eta=\langle W_{a|out}(t)\rangle/\bar{n}_a$, as a function of $\chi t$ for cross-Kerr coupler (continuous red line), 2-photon exchange (blue dashed line) and 3-photon exchange (black dot-dashed line) with input average photon number $\bar{n}_a=1$ and $g=\chi$.  } }
\label{EffVsTCK23}
\end{figure}

\begin{figure}[t]
\begin{center}
\includegraphics[height=6cm,width=7.5cm]{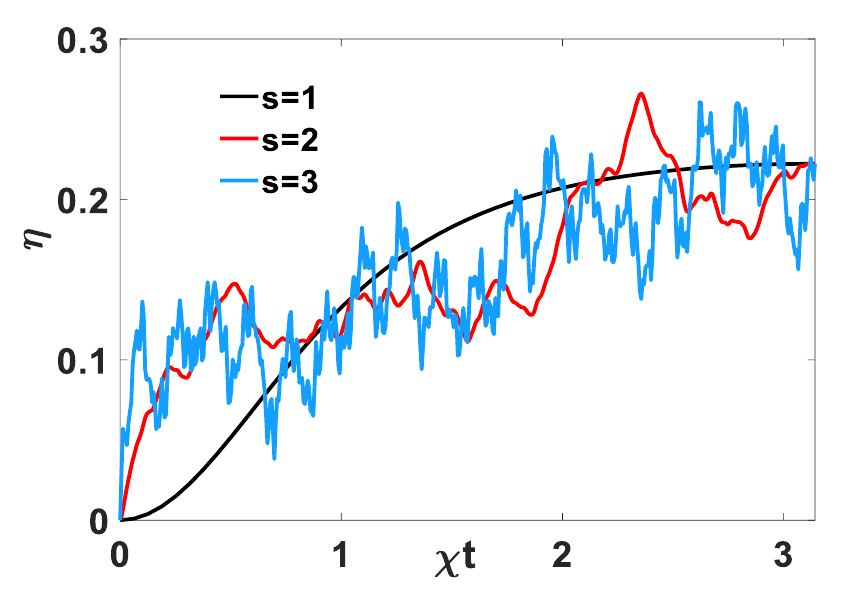}
\end{center}
\caption{{ Efficiency as a function of $\chi t$ for various order of cross-phase couplings for $\bar{n}_a=1$. } }
\label{EffVsThigh}
\end{figure}

\begin{figure}[t]
\begin{center}
\includegraphics[height=6cm,width=7.5cm]{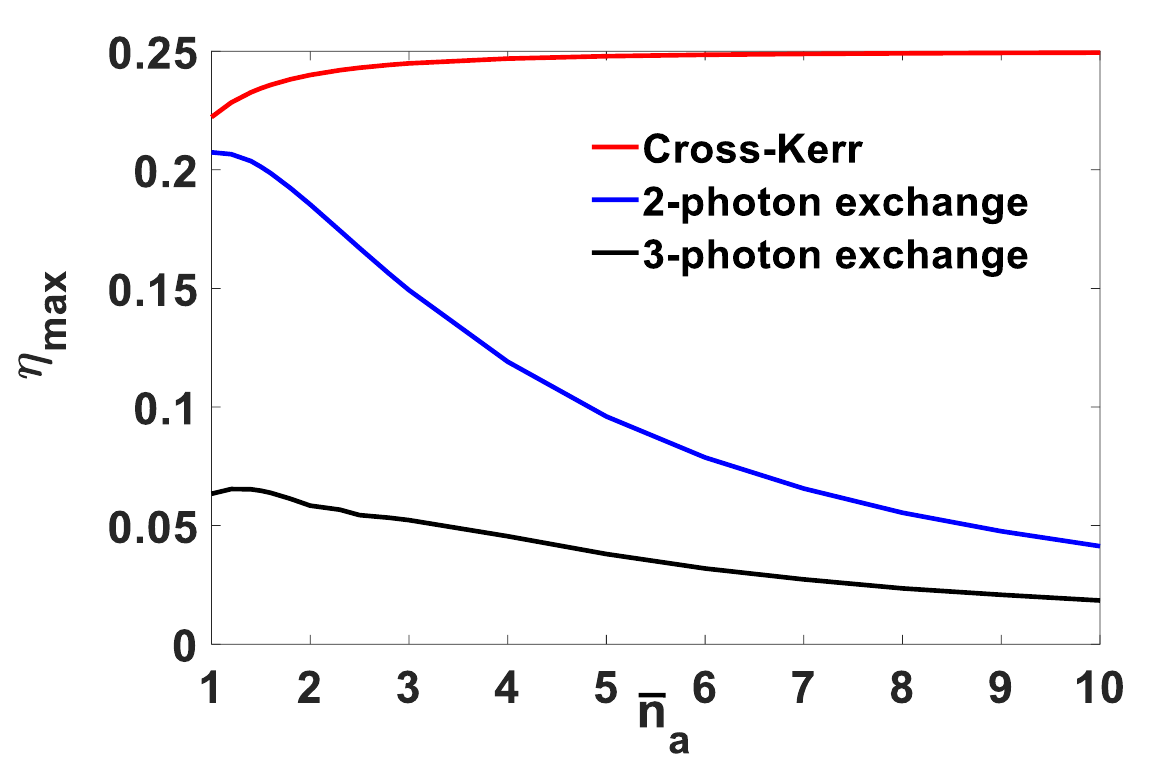}
\end{center}
\caption{Maximal WC efficiency $\eta_{max}=\langle W_{a|out}\rangle_{max}/\bar{n}_a$ of the interferometers with cross-Kerr (red line), two-photon exchange (blue line) and three-photon exchange (black line).  \textit{The saturation behavior is drastically different for each process, therefore can be determined from the output WC.} }
\label{MaxEff}
\end{figure}

Although Eqs. \eqref{RhoAout} and (\ref{ergotropy}) describe both cross-Kerr and even-$k$ photon exchange, their WC varies \textit{quite differently in time}, as shown in Fig. \ref{EffVsTCK23}, and is thus process-specific. \\   
\noindent
(b) \textbf{Odd-$k$ ($k\geq 3$) photon exchange Hamiltonians} do not impose a definite parity on the output state [SI Sec. 2]. Nevertheless, they too yield non-monotonic/non-passive photon number distributions [Fig. \ref{k23PaPb}(b)] with non-zero WC at the output, albeit with distinct interaction-time dependence [Fig. \ref{EffVsTCK23} and Table 1]. The WC is found to \textit{oscillate more rapidly} as $k$ increases, due to the increased rate of energy exchange between the modes.\\  
\noindent
(c) { A salient difference between cross-phase dispersive coupling, such as cross-Kerr, and direct $k$-photon exchange is the \textit{time-dependence} [Figs. \ref{EffVsTCK23} and \ref{EffVsThigh}] and the \textit{maximum value} [Fig. \ref{MaxEff}] of the \textbf{WC efficiency}, i.e., the mean output WC normalized by the mean input energy, that varies between 0 and 1, 
\begin{align}
\eta(t)=\frac{\langle W_{a|out}(t)\rangle}{\bar{n}_a}.
\end{align}  
For thermal input, $P_n=\bar{n}^n_a/(1+\bar{n}_a)^{n+1}$, still keeping the input mode $b$ in the vacuum state, 
the \textit{maximum WC efficiency} of the cross-Kerr process then saturates to $\eta^{CK}_{max}=1/4$ in the classical limit $\bar{n}_a\gg 1$. By contrast, the maximal efficiency for direct photon exchange processes is inversely proportional to the average input photon number in the same limit: For two-photon exchange $(k=2)$, $\eta^{(k=2)}_{max}\approx {0.4}/{\bar{n}_a}$
and for three-photon exchange process $(k=3)$, $\eta^{(k=3)}_{max}\approx {0.2}/{\bar{n}_a}$ [Fig. \ref{MaxEff}}]. Thus, the \textit{output WC efficiency senses the quantum nonlinear process} that the two modes undergo inside the interferometer. {The present analysis, which is geared towards small $n$-numbers, may be extended to higher-order exchange processes if higher $n$-numbers are involved.} 
 
{Parity selection of the output, namely, cancellation of the odd-number states under cross-Kerr/even-k photon exchange requires strong nonlinearity, $\chi t=\pi/2$ in the cross-Kerr case, whereas much weaker nonlinearity does not yield parity selection. The same holds true for processes combining even and odd parity of photon exchange, when the two processes have comparable strength. The resulting WC  may reflect these underlying features, as we have checked numerically.} 
 
To sum up, WC characteristics provide a qualitative signature of the different nonlinear processes for a given thermal input.

\section{Rapport of coherence functions and work capacity}\label{Coherencefunction}
Coherence functions of a quantum field, up to high order, that are commonly recorded by multi-detector coincidence, store full information on the field state \cite{Carmichael_BOOK,Gardiner,
ScullyZubairy}. Here we show that the WC mean and dispersion of a single output mode can invariably convey this information.   

\begin{figure}[t]
\begin{center}
\includegraphics[height=5cm,width=15cm]{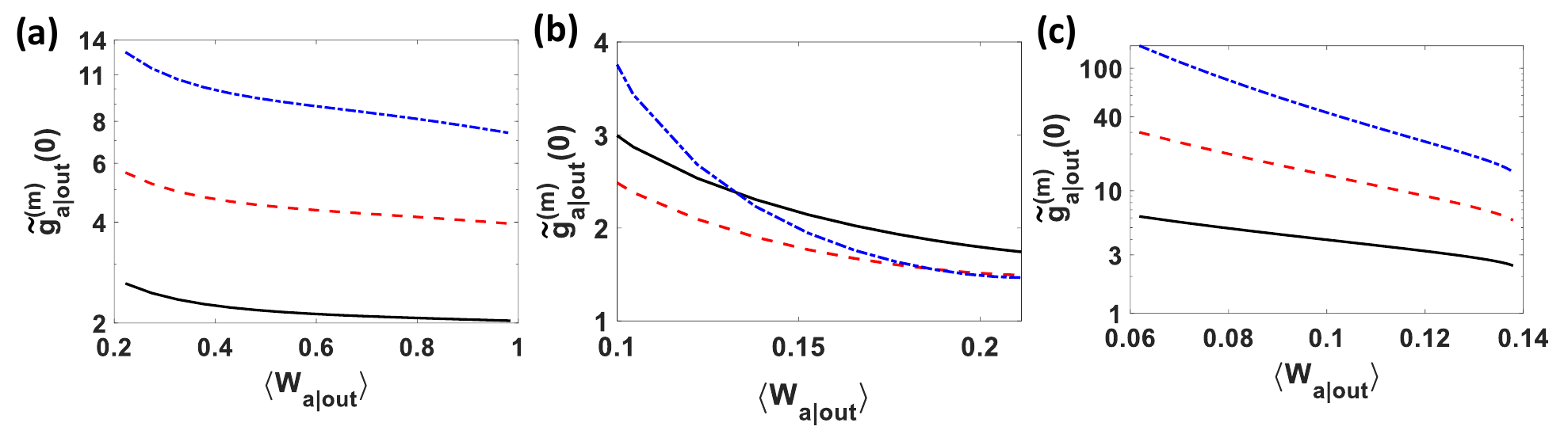}
\end{center}
\caption{ (a) Second- (black solid line), third- (red dashed line) and fourth- (blue dot-dashed line) order coherence functions, normalized by $m!$ ($m$-th order coherence function of the thermal state \cite{Bai2017JOSAB}), of the output mode $a_{out}$ as a function of output WC for cross-Kerr interferometer. (b) same for 2-photon exchange  and (c) 3-photon exchange. These coherence functions can be inferred from the output WC.} 
\label{g2g3g4vschiErgConnk23}
\end{figure}

In the scenario of section, i.e., for thermal state input to mode $a$ and vacuum state input in mode $b$, we may consider coherence functions of the following processes/gauge fields:\\
\noindent
(a) { \textbf{Coherence functions for cross-Kerr or even-$k$ photon exchange :}  We find that the zero time-delay second-order coherence function at the output mode $a$ of these processes can be expressed via the two lowest moments of the WC as follows [SI Sec. 3]
\begin{subequations}\label{g2WDef}
\begin{align}\label{g2W}
g^{(2)}_{a|out}(0)=\frac{\langle \hat a_{out}^{\dagger 2}\hat a_{out}^2\rangle}{\langle \hat a_{out}^{\dagger}\hat a_{out}\rangle ^2}=1-\frac{1}{2\langle W_{a|out}\rangle}+\frac{|\langle \Delta W_{a|out}\rangle^2|}{3 \langle W_{a|out}\rangle^2},
\end{align}
Here we define the WC dispersion as the absolute-value difference between the actual (non-passive-state) output energy variance and its passive-state counterpart
\begin{align}
|\langle \Delta W_{a|out}\rangle^2|=&\left|\left[\langle E_{a|out}^2\rangle  -\langle E_{a|out}\rangle^2\right]\right. \nonumber\\&~~~~ \left. -\left[\langle (E_{a|out}^{pas})^2\rangle-\langle E_{a|out}^{pas}\rangle^2\right]\right|.\end{align}
\end{subequations}}
Relation \eqref{g2W}, which holds for \textit{any average input photon number}, demonstrates that the WC can indeed disclose the coherence function. { The normalized second-, third- and fourth-order coherence functions,
$\tilde{g}^{(m)}_{a|out}(0)=g^{(m)}_{a|out}(0)/g^{(m)}_{th}(0)$, where $g^{(m)}_{th}(0)=m!$ is $m$th-order coherence functions of the thermal state  \cite{Bai2017JOSAB}, are shown as a function of output WC in Fig. \ref{g2g3g4vschiErgConnk23}.} In the limit $\bar{n}_a\ll 1$, for which $\langle W_{a|out}\rangle$ is small, the rapport of WC and coherence functions is particularly transparent: the \textit{coherence functions} $\tilde{g}^{(2)}_{a|out}(0)$ and $\tilde{g}^{(3)}_{a|out}(0)$ are reciprocal, whereas $\tilde{g}^{(4)}_{a|out}(0)$ is quadratically reciprocal, to the WC: 
\begin{subequations}\label{g20W}
\begin{align}
\tilde{g}^{(2)}_{a|out}(0)&\propto \frac{1}{2\langle W_{a|out}\rangle},\\
\tilde{g}^{(3)}_{a|out}(0)&\propto \frac{3 f(t)}{2\langle W_{a|out}\rangle},\\
\tilde{g}^{(4)}_{a|out}(0)&\propto \frac{3 f(t)}{4\langle W_{a|out}\rangle^2},
\end{align}
\end{subequations}
where $f(t)$ is an oscillatory function that depends on $\chi t$ for cross-Kerr MZI and on $gt$ for even-$k$-photon exchange MZI [SI Sec. 3]. However, \textit{even for larger} $\langle W_{a|out}\rangle$ (Fig. \ref{g2g3g4vschiErgConnk23}) there is a clear dependence of $\tilde{g}^{(m)}_{a|out}(0)$ on WC. \\

\noindent
(b) {\textbf{Coherence function for 3-photon exchange: } In this case, as the output state contains both even and odd number of photons, we find that the coherence functions scale with the WC differently [SI Sec. 3]. The scaling is cumbersome but tractable, at least numerically, for large input photon number $\bar{n}_a$. Again, for small $\bar{n}_a$, the scaling becomes transparent,} 
\begin{align}\label{gmout}
\tilde{g}^{(m)}_{a|out}(0)\propto \frac{q_m(t)}{\langle W_{a|out}\rangle^m},~~~(m=2,3,4)
\end{align}  
where $q_m(t)$ express the time dependence specific to the $m$th order.

The expressions in Eqs. (\ref{g2WDef}-\ref{gmout}) indicate that the \textit{multi-order coherence functions at the output can be inferred from the corresponding WC measurements without interferometry (coincidence measurements)}.





\section{WC measurement by a mechanical oscillator}\label{optomechanics}
Work stored in the interferometer output field  can be extracted by a unitary operation, but this operation depends on the output state in a fashion that may not be easily evaluated \cite{PRE2014, OpatrnyPRL21}. {Here, in contrast to Ref. \cite{Opatrny2023ScAdv}, we do not wish to extract work but to infer the WC of the output field.} We show this can be done by its optomechanical coupling to a mechanical oscillator via the Hamiltonian \cite{Law1995PRA,Aspelmeyer2014RevModPhys}
\begin{align}
H_{field+Osc}=\hbar\omega \hat a_{out}^\dagger \hat a_{out}+\hbar \Omega \hat O^\dagger \hat O+\hbar G \hat a_{out}^\dagger \hat a_{out}(\hat O^\dagger+\hat O),
\end{align} 
where $\hat a_{out}^\dagger (\hat a_{out})$ and $\hat O^\dagger (\hat O)$ are the creation (annihilation) operators of the output field and the mechanical oscillator, respectively [Fig. \ref{NLMZI}]; $\omega (\Omega)$ are their corresponding resonance frequencies, and $G$ is the optomechanical coupling strength. 

{Let us initialize the oscillator in a coherent state $\ket{\alpha}$ with average energy $\bar{n}_O=|\alpha|^2$. For an output field prepared by the interferometer containing either cross-Kerr coupling or even-$k$ photon exchange process, for an oscillator initialized in a coherent state,
we can monitor the beating of the oscillator mean energy which is found to satisfy [SI Sec. 4]}
\begin{align}\label{meanphononnumber}
&\langle O^\dagger O\rangle(\tau)= \bar{n}_O+\frac{2\sqrt{2}G}{\Omega}\langle W_{a|out}(t)\rangle\left(\frac{\alpha+\alpha^*}{\sqrt{2}}(1-\cos\Omega \tau)\right. \nonumber\\
&\left.-\frac{\alpha-\alpha^*}{\sqrt{2}i}\sin\Omega \tau\right)+\left[ \frac{|\langle \Delta W_{a|out}(t)\rangle^2|}{3} +\langle W_{a|out}(t)\rangle^2 \right] \frac{16G^2}{\Omega^2}\sin^2\left(\frac{\Omega \tau}{2}\right).
\end{align}
{The beating term that scales with the coherent state amplitude $\alpha$ dominates over the last term on the right hand side provided $|\alpha| \gg |\langle \Delta W_{a|out}(t)\rangle^2|, \langle W_{a|out}(t)\rangle^2$. This term renders the mean oscillator energy-beating proportional to the mean WC $\langle W_{a|out}(t)\rangle$.} 

{The WC time-dependence in Eq. \eqref{meanphononnumber} can be realized by increasing the time duration of the nonlinear interaction, which requires varying the length of the cavity or cell.} {The increase of the nonlinear interaction time by extending the medium length does not conform to a strict definition of a \textquotedblleft black box” , but presumes that all  other characteristics of the Hamiltonian are invariant under such length change. }

Thermal fluctuations of the mechanical oscillator can be negligible compared to the beating term in Eq. \eqref{meanphononnumber} even if the oscillator thermal energy $\bar{n}_{th}$ is non-negligible provided $\bar{n}_O=|\alpha|^2\gg \bar{n}_{th}$.

{Thus, the mechanical-oscillator mean energy, which can be experimentally probed with zero-point accuracy upon recording its \textit{phonon power spectrum} \cite{Chan2011Nature}, provides a simple signature of the underlying quantum nonlinear process, triggered by thermal input to the interferometer.}

\section{Discussion}\label{discussion}
Our scheme and analysis have introduced the hitherto unknown concept of \textit{sensing the quantum fluctuations} of nolinearly correlated two-mode fields via the \textit{lowest moments of a thermodynamic variable, namely, single-mode work capacity (WC)}, alias ergotropy. The fields to be sensed are transformed from thermal input to unknown output by a nonlinear \textquotedblleft black-box" embedded in a Mach-Zehnder interferometer where the output is monitored by a mechanical oscillator. {All  nonlinear  processes investigated by us in this paper, which practically cover the entire range of feasible  two-mode Hamiltonians, have been shown to be characterized by work capacity.}  

{This analysis should be contrasted with our recent exploration of cross-Kerr nonlinear interferometry as a vehicle of \textit{work extraction} from a thermal input, i.e., a nonlinear heat engine \cite{Opatrny2023ScAdv}. The present analysis has another aim: that of distinguishing between different fields generated via nonlinear filtering of thermal noise.}

{This filtering based on the output WC is advantageous in that it \textit{rejects processes} where modes are linearly coupled (e.g. via Raman interaction) and thus yields Gaussian output void of WC. By contrast, quadratic, cubic or quartic photon number interactions i.e, 4th, 6th, 8th order interactions in the field operators yield unique output WC (all higher-order exchange processes are negligible).}

{The proposed sensing is simpler than quantum process tomography (QPT) of the two-mode output, which is so demanding that it is rarely used \cite{Lvovsky2009RMP}. In QPT of two-mode continuous variables, the fidelity would scale with the number of measurements of both modes, which may require staggering and unmanageable numbers of measurements.}

{By contrast, in the proposed method, the discrimination between different nonlinear gauge fields has been shown to require the much simpler recording of the oscillator power spectrum that discloses the oscillator mean energy \cite{Chan2011Nature} and thereby the single-mode output mean WC. A major extension of the existing quantum thermodynamic resource theory beyond linear Gaussian operations \cite{Horodecki2013NatComm,Brandao2015PNAS,Narasimhachar2021NPJQI,Serafini2020PRL} to operations on nonlinear continuous variables, may allow a quantitative comparison with our method.}

{We have focused on autonomous unitary two-mode input to two-mode output transformations. One may similarly diagnose non-autonomous (pump-induced) three-wave mixing transformations \cite{boyd,Jouravlev2004PRA,drummond2014quantum,Ghosh2017} of the thermal input such as parametric down-conversion and second harmonic generation [SI Sec. 2] as well as non-unitary (measurement induced) photon addition or subtraction to the thermal input \cite{Opatrny2000PRA,OpatrnyPRL21}.} 

The present noise sensing scheme is experimentally feasible and can diagnose diverse multi-photon processes observable in: 1) atomic or molecular media enclosed in high-Q cavities \cite{Kurizki2000NJP} or microspheres \cite{Jouravlev2004PRA}, 2) charged-particle interactions in solids \cite{Gevorkyan2015LPL,Kurizki1988PRA,Kurizki1985PRA} that may generate short-wavelength photon pairs, 3) cross-Kerr coupling at the level of few photons in cold-atom high density media \cite{Opatrny2023ScAdv} or in circuit QED \cite{Kounalakis2018npjQI,Jin2013PRL}.  State-of-the art optomechanical setups \cite{Aspelmeyer2014RevModPhys, Treutlein2014NatNanotech,Chan2011Nature, Schliesser2009NatPhys, Bild2023Science} allow high-precision monitoring of the mean energy of a mechanical oscillator, and therefore, as shown here, can provide reliable work capacity measurements.

{Notwithstanding the recent demonstrations of giant Kerr-nonlinearity in cold-polariton media \cite{drori2023quantum} and in  high-Q cavities \cite{kirchmair2013observation, vrajitoarea2020quantum}  and the possibility of implementing multi-order photon exchange in these systems,  the present “black-box” approach can only become viable when  such  processes are realized in many diverse systems, whose complexity challenges quantum tomography. The characterization of these processes via the  work capacity of an output mode, probed by its optomechanical coupling to a driven oscillator, relies on recent advances in quantum optomechanics \cite{Aspelmeyer2014RevModPhys, Treutlein2014NatNanotech,Chan2011Nature, Schliesser2009NatPhys, Bild2023Science} but requires the development of dedicated methods, just as any new technique.}

The proposed sensing may find intriguing applications in diverse quantum technologies \cite{Kurizki2015PNAS,Spasibko2017PRL, Berchera2019Metrologia,
Slussarenko2019AppPhysRev} where the knowledge of single-mode quantum statistics and two-mode correlations is important: Either super-bunching  and super-thermal photon statistics, which have a growing scope of  applications \cite{Hong2019AppSc,Spasibko2017PRL}, or, conversely, antibunching and sub-Poissonian photon statistics,  which are almost indispensable for  quantum communication and quantum information processing \cite{Davidovich1996RevModPhys,Lounis2005RepProgPhys, Kiraz2004PRA}. {Quantum electrodynamic gauge-field \cite{Goldman2014PRX, delPino2023PRL} generation is another emerging area that may appreciably benefit from this sensing }. 

To conclude, the conceptual advantages of the procedure may become a viable alternative to process tomography of interferometric measurements \cite{Chuang1997JModOpt, Arino2003AdvImagElecPhys,Mohseni2008PRA}.

\noindent
\textbf{Acknowledgements}\\ 
NM acknowledges the support of the Feinberg Graduate School (FGS) Dean Postdoctoral Fellowship. GK is supported by DFG (FOR 7024), EC (PATHOS, FET Open), Quantera (PACE-IN), and the US-Israel NSF-BSF. TO was supported by the Czech Science Foundation, grant 20-27994S.

\renewcommand{\thefigure}{S\arabic{figure}}
\renewcommand{\theequation}{S\arabic{equation}} 
\setcounter{equation}{0}
\setcounter{figure}{0}
\setcounter{section}{0}

\section*{Supplementary Information}

\section{Nonlinear Hamiltonians in terms of Stokes operators}\label{NonlinearHamiltonian}
The gauge-field Hamiltonians that we discuss are nonlinear functions of two-mode Stokes operators, or equivalently, of pseudospin operators $J_x=(\hat a^\dagger \hat b+\hat a \hat b^\dagger)/2$, $J_y=(\hat a^\dagger \hat b-\hat a \hat b^\dagger)/2i$, and $J_z=(\hat a^\dagger \hat a - \hat b^\dagger \hat b)/2$; where $\hat a$ and $\hat b$ are the annihilation operators of the two coupled modes.
\subsection{Cross-Kerr coupling}
In terms of Stokes operators
\begin{align}
\hat n_a \hat n_b=\frac{1}{4}\hat N^2-\hat J_z^2.
\end{align} 
The cross-Kerr interaction corresponds to \textquotedblleft one-axis twisting" model of spin squeezing on the Bloch sphere \cite{Kitagawa1993PRA}, where the dynamics have two stable stationary points on the poles, $\langle \hat J_x\rangle=\langle \hat J_y\rangle=0, \langle \hat J_z\rangle=\pm\langle \hat N\rangle/2$ and a circle of unstable stationary points at equator.  Starting with a spin coherent state and placing it at unstable stationary points leads to the fastest squeezing \cite{Opatrny2015PRA, Opatrny2015PRA2}.   \\ 
\subsection{Multi-photon exchange}
The multi-photon exchange interaction Hamiltonian in terms of Stokes operators is
\begin{align}
\hat a^{\dagger k} \hat b^k+\hat a^k \hat b^{\dagger k}=(\hat J_x+i\hat J_y)^k+(\hat J_x-i\hat J_y)^k.
\end{align}
For $k=2$, 
\begin{align}
\hat a^{\dagger 2} \hat b^2+\hat a^2 \hat b^{\dagger 2}=2(\hat J_x^2-\hat J_y^2),
\end{align}
corresponds to \textquotedblleft two-axis countertwisting" model of spin squeezing \cite{Kitagawa1993PRA}. The dynamics has four stable points at  $\langle \hat J_x\rangle=\pm\langle \hat N\rangle/2$ and at $\langle \hat J_y\rangle=\pm\langle \hat N\rangle/2$, and two unstable stationary points at the poles $\langle J_z\rangle=\pm\langle \hat N\rangle/2$ which are the optimum starting points for generating spin squeezed states. 

For $k=3$, 
\begin{align}
\hat a^{\dagger 3} \hat b^3+\hat a^3 \hat b^{\dagger 3}=2(\hat J_x^3-\hat J_x\hat J_y^2-\hat J_y\hat J_x\hat J_y-\hat J_y^2\hat J_x).
\end{align}

\section{Work capacity (Ergotropy) in nonlinear interferometry}\label{ErgotropyWCSI}
\subsection{Basic principles}
The  mean work capacity (WC) or ergotropy \cite{kurizki2022Book,PRE2014,Ghosh2018,
 niedenzu18quantum,Allahverdyan_2000} $\langle W_{a|out}\rangle$ of one output mode, labeled by $a$, is given by
\begin{subequations}\label{WC}
\begin{align}
\langle W_{a|out}\rangle&=\langle E_{a|out}\rangle-\langle E_{a|out}^{pas}\rangle,\\
E_{a|out}^{pas}&=\text{Tr}(\hat U_{perm}\rho_{a|out}\hat U_{perm}^\dagger \hat H_a).
\end{align}
\end{subequations}
Here, $\langle E_{a|out}\rangle=\text{Tr}(\hat H_a\rho_{a|out})$, $\hat H_a=(\hat a^\dagger \hat a+1/2)\hbar\omega$  and $\rho_{a|out}=\text{Tr}_b(\hat U \rho_{in}\hat U^\dagger)$ is the reduced density matrix of the output mode $a_{out}$. The mean non-passive energy is the difference between the total energy $\langle E_{a|out}\rangle$ and the passive part. The passive part is obtained by applying a unitary permutation transformation $\hat U_{perm}$, which arranges the output probabilities in descending order of energy, thus resulting in a passive state \cite{kurizki2022Book,Allahverdyan_2000,PRE2014,Ghosh2018,
 niedenzu18quantum}.

{We define the WC dispersion as
\begin{align}
|\langle \Delta W_{a|out}\rangle^2|=\left|\left[\langle (a_{out}^\dagger a_{out})^2\rangle-\langle a_{out}^\dagger a_{out}\rangle^2\right]-\left[\langle (a_{out}^\dagger a_{out})^2\rangle_{pas}-\langle a_{out}^\dagger a_{out}\rangle_{pas}^2\right]\right|,
\end{align}
where the expectation values $\langle (a_{out}^\dagger a_{out})^2\rangle$ and $\langle a_{out}^\dagger a_{out}\rangle$ are to be calculated in the output state $\rho_{a|out}$, and  $\langle (a_{out}^\dagger a_{out})^2\rangle_{pas}$ and $\langle a_{out}^\dagger a_{out}\rangle_{pas}$ are to be calculated in the passive state $\rho_{a|out}^{pas}$.}
\subsection{WC in Cross-Kerr Mach-Zehnder interferometer (MZI)}
To show that single-mode WC can be generated from any passive state that is filtered through a cross-Kerr MZI, consider the input to mode $a$ in a passive state,
\begin{align}\label{InputA}
\rho_{a}=\sum_{n}P_n\ket{n}\bra{n},
\end{align}
where $P_n\geq P_{n+1}$, while mode $b$ in vacuum state. {The output state is
\begin{align}\label{Rhoout}
\rho_{out}=\hat U (\rho_{a} \otimes \ket{0}\bra{0})\hat U^\dagger,
\end{align}
where
\begin{align}\label{UnitaryNLMZICK}
 \hat U=\hat U_{BS}\hat U_{NL}(\chi t)\hat U_{BS},
\end{align} 
 $\hat U_{BS}$ and $\hat U_{NL}(\chi t)$ being the beam splitter and nonlinear cross-Kerr transformations respectively.    
We rewrite Eq. \eqref{UnitaryNLMZICK} as
\begin{align}
\hat U=\hat U_{BS} e^{-i\chi t \hat n_a \hat n_b}\hat U_{BS}&=\hat U_{BS} e^{-i\chi t \hat n_a \hat n_b}\hat U_{BS}^\dagger\hat U_{BS}\hat U_{BS}\nonumber\\
&=\hat U_{BS} (1-i\chi t \hat n_a \hat n_b+\cdots ) \hat U_{BS}^\dagger\hat U_{BS}\hat U_{BS}.
\end{align}
The transformation of the $\chi t$-order term 
\begin{align}
\hat U_{BS} \hat n_a \hat n_b \hat U_{BS}^\dagger= \frac{1}{4} \left[\hat a^{\dagger 2} \hat a^2 +\hat b^{\dagger 2} \hat b^2 + \hat a^{\dagger 2} \hat b^2 + \hat a^2 \hat b^{\dagger 2}\right],
\end{align}
generates the two-photon transition along with self-Kerr terms. Therefore, the cross-Kerr nonlinearity sandwiched between the two beam splitters is equivalent to two-photon transition, and yields even-$n$ exchange. Therefore, upon tracing the output mode $b$ from the state $\rho_{out}$, we get the output state of the mode $a$ to be
\begin{align}\label{RhoAoutSI}
\rho_{a|out}=\tilde{P}_0\ket{0}\bra{0}+\sum_{n~even}\tilde{P}_n \ket{n}\bra{n},
\end{align}
in which the odd number states are missing. Thus, the probability distribution becomes non-monotonic and stores ergotropy. }

The output average photon number is
\begin{align}\label{outputAdA}
\langle a_{out}^\dagger a_{out}\rangle=\sum_{n~even} n\tilde{P}_n.
\end{align}

The passive state that can be constructed from the state $\rho_{a|out}$ (in Eq. \eqref{RhoAoutSI}) is
\begin{align}\label{RhoAPas}
\rho_{a|out}^{pas}=\tilde{P}_0\ket{0}\bra{0}+\sum_{m=1}^{\infty} \tilde{P}_{2m} \ket{m}\bra{m},
\end{align}
where $n=2m$ is an even number, and its passive energy is
\begin{align}\label{passiveAdA}
\langle a_{out}^\dagger a_{out}\rangle_{pas}=\sum_{m=1}^{\infty} m\tilde{P}_{2m}=\frac{\langle a_{out}^\dagger a_{out}\rangle}{2}.
\end{align}
The WC is
\begin{align}\label{WCExp}
\langle W_{a|out} \rangle=\langle a_{out}^\dagger a_{out}\rangle-\langle a_{out}^\dagger a_{out}\rangle_{pas}=\frac{\langle a_{out}^\dagger a_{out}\rangle}{2}=\frac{\sum_{n~even} n\tilde{P}_n}{2}.
\end{align}




For thermal input,
the above equation yields
\begin{align}\label{WCthermal}
\langle W_{a|out}\rangle= \frac{\langle a_{out}^\dagger a_{out}\rangle}{2}=\frac{\bar{n}_a}{4}\left(1-\frac{1}{(1+\bar{n}_a-\bar{n}_a\cos\chi t)^2} \right).
\end{align}


For this interferometer, from Eqs. \eqref{RhoAoutSI}, \eqref{RhoAPas} and \eqref{passiveAdA}, we find that
\begin{align}
\langle (a_{out}^\dagger a_{out})^2\rangle &=4 \langle (a_{out}^\dagger a_{out})^2\rangle_{pas},\\
\langle a_{out}^\dagger a_{out}\rangle &=2 \langle a_{out}^\dagger a_{out}\rangle_{pas},
\end{align}
which yield the expression
\begin{align}\label{VarErg}
|\langle \Delta W_{a|out}\rangle^2|=\frac{3}{4} (\langle (a_{out}^\dagger a_{out})^2\rangle-\langle a_{out}^\dagger a_{out}\rangle^2),
\end{align}
that is related to the number fluctuation of the output mode $a$.

We have checked that the output state of the interferometer containing  high-order cross-phase coupling with unitary operator $U_{NL}^{(s)}=\exp[-i\chi t \hat n_a^s  \hat n_b^s]$ is also of the form given in Eq. \eqref{RhoAoutSI}. Therefore, WC satisfies the relation given in Eq. \eqref{WCExp}. However, the time-dependence of WC is different for different order of couplings, as can be seen in Fig. \ref{EffVsTCK23}.


\subsection{WC in two-photon exchange MZI}
Now, consider the nonlinear interaction to be 2-photon exchange process. Then the total unitary operator becomes
\begin{align}
\hat U=\hat U_{BS}\hat U_{NL}^{(2)}\hat U_{BS},
\end{align}
where $U_{NL}^{(2)}=e^{-ig(a^{\dagger 2} b^2+a^2 b^{\dagger 2})t}$ is a two-photon exchange process.

{
The MZI transformation of the two-photon exchange Hamiltonian 
\begin{align}
\hat U_{BS} (\hat a^{\dagger 2} \hat b^2 + \hat a^2 \hat b^{\dagger 2}) \hat U_{BS}^\dagger = \frac{1}{2} \left[-\hat a^{\dagger 2} \hat a^2 -\hat b^{\dagger 2} \hat b^2 + \hat a^{\dagger 2} \hat b^2 + \hat a^2 \hat b^{\dagger 2}+4 \hat n_a \hat n_b\right],
\end{align}
generates a two-photon transition along with cross-Kerr and self-Kerr terms. }

\begin{figure}
\begin{center}
\includegraphics[height=10cm,width=15cm]{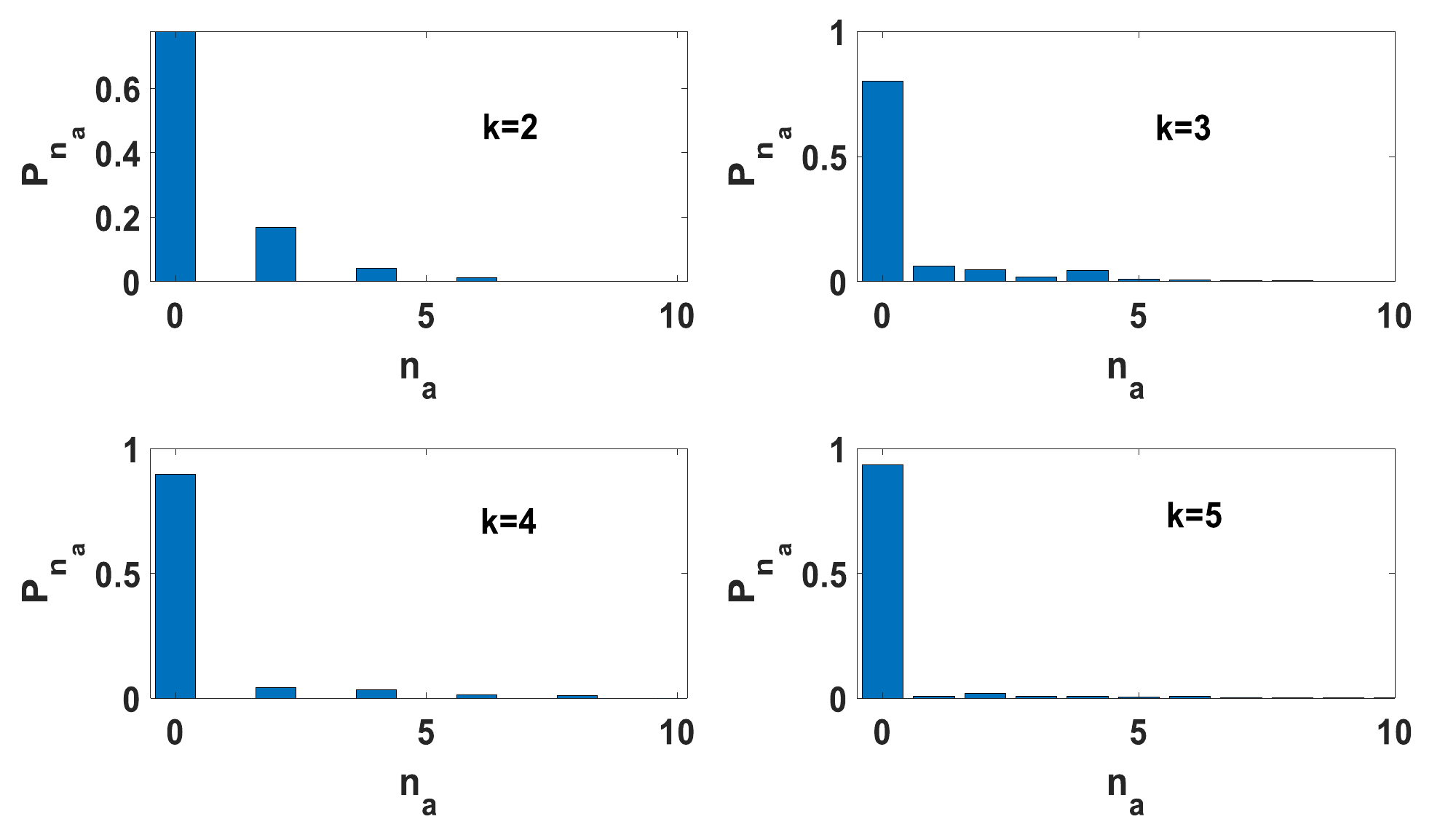}
\end{center}
\caption{Photon number distribution at the output mode $a_{out}$ for $k$-photon exchange processes with $k=2,3,4,5$. For even-$k$ photon exchange processes, the output state contains only even number of photons. }\label{ProbabilityDistk2345}
\end{figure}

Consider the input passive state $\rho_{a}=\sum_{n}P_n\ket{n}\bra{n}$ and the mode $b$ in vacuum. Then the time-evolved state of mode $a$ after tracing out the mode $b$ is
\begin{align}
\rho_{a|out}(t)=\text{Tr}_b(\rho(t))=\tilde{P}_0\ket{0}\bra{0}+\sum_{n~even} \tilde{P}_n(t)\ket{n}\bra{n},
\end{align}
where\\
\begin{subequations}\label{LeadingProbabilitiesk2}
\begin{align}
\tilde{P}_0(t)&\approx P_0+P_1+P_2 \cos^2 gt+P_3\left(\frac{3}{4}\sin^2 2\sqrt{3}gt+\cos^2 2\sqrt{3}gt\right)...,\\
\tilde{P}_2(t)&\approx P_2 \sin^2 gt+P_3\frac{1}{4}\sin^2 2\sqrt{3}gt+P_4\frac{1}{8} \sin^2 4\sqrt{3}gt+...,\\
\tilde{P}_4(t)&\approx P_4\frac{1}{16} \left|2\cos 4\sqrt{3}gt+i\sqrt{3} \sin 4\sqrt{3}gt-2\cos 6gt-2i\sin 6gt\right|^2+...,
\end{align}
\end{subequations}
The output state contains only even number of photons. This trend also arises for higher even-$k$ photon exchange processes [see Fig. \ref{ProbabilityDistk2345}].

As the output state contains only even number of photons, 
the WC for this state is
\begin{align}\label{Ergevenk}
\langle W_{a|out}\rangle= \frac{\langle a^\dagger_{out}a_{out}\rangle}{2}=\frac{\sum_{n~even}\tilde{P}_n}{2},
\end{align}
and the WC dispersion is
\begin{align}\label{VarErgKeven}
\left|\langle \Delta W_{a|out}\rangle^2\right|=\frac{3}{4} (\langle (a_{out}^\dagger a_{out})^2\rangle-\langle a_{out}^\dagger a_{out}\rangle^2),
\end{align}
which have similar expressions to that of the cross-Kerr MZI.

\subsection{WC in three-photon exchange MZI}
Consider the three-photon exchange process between the modes, defined by the unitary operator $U_{NL}^{(3)}=e^{-ig(a^{\dagger 3} b^3+a^3 b^{\dagger 3})t}$. 

{The MZI transformation of three-photon exchange Hamiltonian 
\begin{align}
\hat U_{BS} (\hat a^{\dagger 3} \hat b^3 + \hat a^3 \hat b^{\dagger 3}) \hat U_{BS}^\dagger=&\frac{1}{4}\left[a^{\dagger 3} \hat b^3 + \hat a^3 \hat b^{\dagger 3}+ 9( \hat a^{\dagger 2} \hat a \hat b^\dagger \hat b^2+\hat a^{\dagger } \hat a^2 \hat b^{\dagger 2} \hat b)\right.\nonumber\\
&\left.-6(\hat a^{\dagger 3} \hat a^2 \hat b+\hat a^{\dagger 2} \hat a^3 \hat b^\dagger+\hat a^{\dagger } \hat b^{\dagger 2} \hat b^3+\hat a \hat b^{\dagger 3} \hat b^2)\right],
\end{align}
which does not give rise to parity-selective transitions.}

For the same input passive state, we get the evolved state of mode $a$ after tracing out the mode $b$ to be
\begin{align}
\rho_{a|out}(t)=\text{Tr}_b(\rho(t))=\sum_n \tilde{P}_n(t)\ket{n}\bra{n},
\end{align}
where
\begin{subequations}\label{LeadingProbabilitiesk3}
\begin{align}
\tilde{P}_0(t)&\approx P_0+P_1+P_2+\left|P_3\left(\frac{1}{2}+\frac{e^{-3igt}}{2}\cos 3gt\right)\right|^2...\\
\tilde{P}_1(t)&\approx \frac{3}{16}P_3\sin^2 6gt+\frac{9}{16}P_4 \sin^2 12gt+...,\\
\tilde{P}_2(t)&\approx \frac{3}{4}P_3  \sin^4 3gt+\frac{3}{8}P_4 \sin^4 6gt+...,\\
\tilde{P}_3(t)&\approx \frac{1}{16}P_3 \sin^2 6gt+\frac{1}{16}P_4 \sin^2 12gt+...\\
\tilde{P}_4(t)&\approx \frac{9}{16}P_4 \sin^4 6gt+...
\end{align}
\end{subequations}
As the output state contains both even and odd number of photons, the output WC does not satisfy Eq. \eqref{WCExp}. For small input average photon number, the WC is generated at the output either if $\tilde{P}_2(t)>\tilde{P}_1(t)$ or $\tilde{P}_3(t)>\tilde{P}_2(t)$. Thus, the WC is
\begin{subequations} \label{WCk3}
\begin{align}
\langle W_{a|out}\rangle &\approx \tilde{P}_2(t)-\tilde{P}_1(t)\approx P_3\left(\frac{3}{4} \sin^{4} 3gt-\frac{3}{16} \sin^2 6 gt\right),\\
&\text{for $(4j+1)\pi/12 <gt<(4j+3)\pi/12$,} \nonumber\\
\langle W_{a|out}\rangle &\approx \tilde{P}_3(t)-\tilde{P}_2(t) \approx P_3\left(\frac{1}{16} \sin^2 6 gt-\frac{3}{4} \sin^{4} 3gt\right),\\
&\text{for $(6j+5)\pi/18 <gt<(6j+7)\pi/18$,} \nonumber
\end{align}
\end{subequations}
where $j=0,1,2..$ with $P_n=\bar{n}_a^n/(1+\bar{n}_a)^{(n+1)}$ is the thermal probability distribution.

\begin{table*}
\caption{Mean WC $\langle W_{a|out}(t)\rangle$ for thermal input distribution $P_n=\frac{\bar{n}_a^n}{(1+\bar{n}_a)^{(n+1)}}$.}\label{table}
\begin{tabular}{|c|m{10cm}|}
\hline
Cross-Kerr & $
\langle W_{a|out}(t)\rangle= \frac{\bar{n}_a}{4}\left(1-\frac{1}{(1+\bar{n}_a-\bar{n}_a\cos\chi t)^2} \right)$.   \\
Cross-Kerr ($\bar{n}_a\ll 1$) &  $\langle W_{a|out}(t)\rangle \approx \frac{\bar{n}_a^2}{(1+\bar{n}_a)^{3}} \sin^2\left(\frac{\chi t}{2}\right)$\\
\hline
2-photon exchange ($\bar{n}_a\ll 1$)  & $\langle W_{a|out} (t)\rangle \approx \frac{\bar{n}_a^2}{(1+\bar{n}_a)^{3}}\left(\sin^{2} gt+\frac{\bar{n}_a}{4(1+\bar{n}_a)} \sin^2 2\sqrt{3} gt \right)$.\\
\hline
3-photon exchange ($\bar{n}_a\ll 1$) & $\langle W_{a|out}(t)\rangle \approx \frac{\bar{n}_a^3}{(1+\bar{n}_a)^{4}}\left(c(t) \sin^{4} 3gt+d(t) \sin^2 6 gt\right).$\\
\hline
\end{tabular}
\end{table*}

\begin{figure}
\begin{center}
\includegraphics[height=10cm,width=15cm]{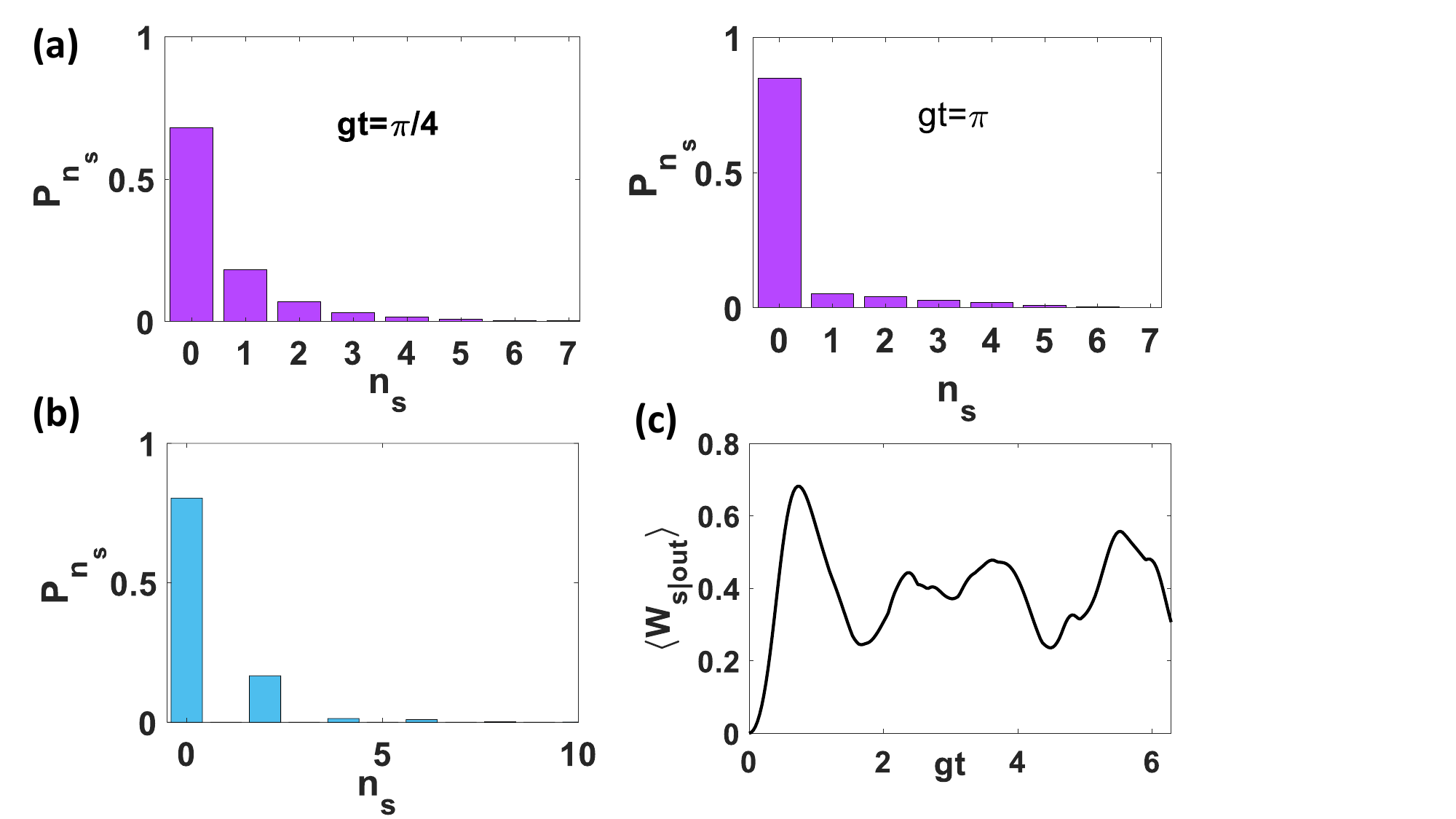}
\end{center}
\caption{Photon number distribution in the signal-mode for non-degenerate parametric down-conversion process with (a) $gt=\pi/4$ (left) and $\pi$ (right). (c) Photon number distribution in the signal-mode for degenerate parametric down-conversion process with $gt=\pi/2$. (d) The output WC in the signal-mode as a function of $gt$ for degenerate parametric down-conversion process. In all plots, the input is thermal state with  $\bar{n}_a=1$. }\label{SPDC}
\end{figure}

\subsection{WC in parametric processes}
{The Hamiltonian that describes the non-degenerate parametric down-conversion process (PDC) is \cite{Drobny1993PRA, McNeil1983PRA,drummond2014quantum,Meher2020JOSAB}
\begin{align}
H_{PDC}=g(a_p a_s^\dagger a_i^\dagger +a_p^\dagger a_s a_i),
\end{align}
where the operators $a_p (a_p^\dagger), a_s (a_s^\dagger)$ and $a_i (a_i^\dagger)$ are the annihilation (creation) operators for the pump, signal and idler fields, respectively; and $g$ is the coupling constant related to the second-order nonlinear susceptibility \cite{boyd,drummond2014quantum,Schneeloch2019JOpt,Meher2020JOSAB}. }

{The photon number distribution of the output signal-mode is shown in Fig. \ref{SPDC}(a) for $gt=\pi/4$ and $\pi$ by considering the input pump-mode in a thermal state. As seen from the figure, the output photon number distributions monotonically fall and thus, the process does not render non-passive output state with non-zero WC. The above Hamiltonian also describes the non-degenerate second harmonic generation (SHG) process and thus, it does not generate WC at the output.}

{The Hamiltonian for the degenerate PDC process is \cite{drummond2014quantum,Drobny1993PRA,Meher2020JOSAB} 
\begin{align}
\tilde{H}_{PDC}=g(a_p a_s^{\dagger 2} +a_p^\dagger a_s^2).
\end{align}
The output photon number distribution becomes non-monotonic [see Fig. \ref{SPDC}(b)]  and therefore, the output signal-mode has non-zero WC [see Fig. \ref{SPDC}(c)]. The above Hamiltonian also describes the degenerate SHG process and thus, generates WC at the output.}

\section{Rapport of coherence functions with WC}\label{CoherenceFunctionSI}

\subsection{Cross-Kerr MZI}
From Eqs. \eqref{WCExp} and \eqref{VarErg}, we find that the zero time-delay second-order coherence function is related to the WC and its dispersion through
\begin{align}\label{g2ErgConnection}
g^{(2)}_{a|out}(0)=\frac{\langle a_{out}^{\dagger 2} a^2_{out}\rangle}{\langle a_{out}^\dagger a_{out}\rangle^2}=1-\frac{1}{2\langle W_{a|out}\rangle}+\frac{|\langle \Delta W_{a|out}\rangle^2|}{3\langle W_{a|out}\rangle^2}.
\end{align} 
For small input average photon number $\bar{n}_a\ll 1$, $|\langle \Delta W_{a|out}\rangle^2| \approx 3\tilde{P}_2-3\tilde{P}_2^2\approx 3\langle W_{a|out}\rangle-3\langle W_{a|out}\rangle^2$ and Eq. (\ref{g2ErgConnection}) yields
\begin{align}
g^{(2)}_{a|out}(0)\approx\frac{1}{2\langle W_{a|out}\rangle},
\end{align}
which indicates that the second-order coherence function at the output can be inferred only from output WC.

To relate the WC with higher-order coherence functions, a few leading output probabilities in Eq. \eqref{RhoAoutSI} for small $\bar{n}_a$ are calculated to be
\begin{subequations}\label{LeadingProbabilities}
\begin{align}
\tilde{P}_2&\approx P_2\sin^2\left(\frac{\chi t}{2}\right),\\
\tilde{P}_4&\approx \frac{P_4}{2}(7+8\cos\chi t+3\cos 2\chi t)\sin^4\left(\frac{\chi t}{2}\right),
\end{align}
\end{subequations}
where $P_n={\bar{n}_a^n}{(\bar{n}_a+1)^{-(n+1)}}$ is the thermal probability distribution. 

The third- and fourth-order coherence functions are, from Eqs. \eqref{RhoAoutSI} and (\ref{LeadingProbabilities}),
\begin{align}
g^{(3)}_{a|out}(0)&=\frac{\langle a_{out}^{\dagger 3} a^3_{out}\rangle}{\langle a_{out}^\dagger a_{out}\rangle^3}\approx \frac {24 \tilde{P}_4}{(2\tilde{P}_2)^3}\approx  \frac{3f(t)}{2\langle W_{a|out}\rangle},\\ 
g^{(4)}_{a|out}(0)&=\frac{\langle a_{out}^{\dagger 4} a^4_{out}\rangle}{\langle a_{out}^\dagger a_{out}\rangle^4}\approx \frac {24 \tilde{P}_4}{(2\tilde{P}_2)^4}\approx \frac{3f(t)}{4\langle W_{a|out}\rangle^2},
\end{align}
where $f(t)=(7+8\cos\chi t+3\cos 2\chi t)$, and we used $\tilde{P}_4\approx P_2^2$ for thermal distribution with $\bar{n}_a\ll 1$. \\

\noindent
\subsection{High-order cross-phase coupling} 
\begin{figure}
\begin{center}
\includegraphics[height=10.5cm,width=15cm]{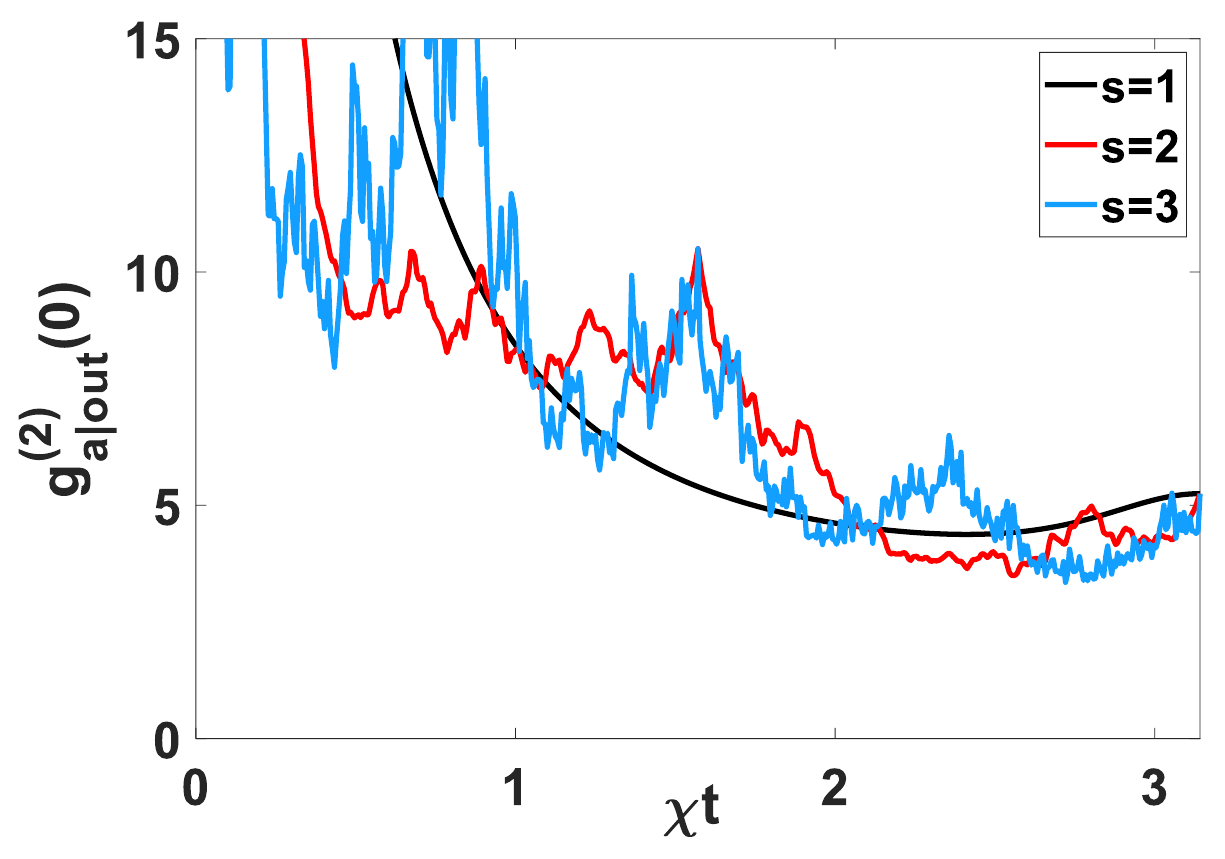}
\end{center}
\caption{ $g^{(2)}_{a|out}(0)$ as a function of $\chi t$ for various order of cross-phase coupling for the input mean photon number $\bar{n}_a=1$.  }\label{g2CrossKerr123}
\end{figure}

The zero time-delay second-order coherence functions for the output field from the interferometer containing various high-order cross-phase couplings are plotted as a function of $\chi t$ in Fig. \ref{g2CrossKerr123}.

\subsection{2-photon exchange MZI}
 From Eqs. \eqref{Ergevenk} and \eqref{VarErgKeven}, we find a similar relation to that of the cross-Kerr MZI, that is, 
\begin{align}\label{g2ErgConnectionkeven}
g^{(2)}_{a|out}(0)=1-\frac{1}{2\langle W_{a|out}\rangle}+\frac{|\langle \Delta W_{a|out}\rangle^2|}{3\langle W_{a|out}\rangle^2},
\end{align}
and for small average photon number input, 
\begin{align}
g^{(2)}_{a|out}(0)&\approx  \frac{1}{2\langle W_{a|out}\rangle}.
\end{align}

From Eqs. \eqref{LeadingProbabilitiesk2}, the higher-order coherence functions are
\begin{align}
g^{(3)}_{a|out}(0)&=\frac{\langle a_{out}^{\dagger 3} a^3_{out}\rangle}{\langle a_{out}^\dagger a_{out}\rangle^3}\approx \frac {24 \tilde{P}_4}{(2\tilde{P}_2)^3}\approx \frac{3f(t)}{2 \langle W_{a|out}\rangle},\\ 
g^{(4)}_{a|out}(0)&=\frac{\langle a_{out}^{\dagger 4} a^4_{out}\rangle}{\langle a_{out}^\dagger a_{out}\rangle^4}\approx \frac {24 \tilde{P}_4}{(2\tilde{P}_2)^4}\approx \frac{3f(t)}{4 \langle W_{a|out} \rangle^2} ,
\end{align}
where
\begin{align}
f(t)=\frac{\left(15+\cos 8\sqrt{3}gt-16\cos gt \cos 4\sqrt{3} gt-8\sqrt{3}\sin 6 gt \sin 4\sqrt{3} gt\right)}{16\sin^4 gt}.
\end{align}

\subsection{3-photon exchange MZI:} 
For $(4j+1)\pi/12 <gt<(4j+3)\pi/12$, using Eqs. \ref{WCk3}(a-b) and Eqs. \ref{LeadingProbabilitiesk3}(a-e),  we find
\begin{align}
g^{(2)}_{a|out}(0)&\approx \frac{2 \tilde{P}_2+6\tilde{P}_3}{(3\tilde{P}_3+2\tilde{P}_2+\tilde{P}_1)^2} \approx  \frac{1}{\langle W_{a|out}\rangle} \frac{(\frac{3}{2} \sin^4 3gt+\frac{6}{16}\sin^2 6gt) \left(\frac{3}{4} \sin^{4} 3gt-\frac{3}{16} \sin^2 6 gt\right) }{\left(\frac{3}{16}\sin^2 6gt+\frac{3}{2} \sin^{4} 3gt+\frac{3}{16} \sin^2 6 gt\right)^2}\approx \frac{q_1(t)}{\langle W_{a|out}\rangle},\\
g^{(3)}_{a|out}(0)&\approx \frac{6\tilde{P}_3+24 \tilde{P}_4}{(4\tilde{P}_4+3\tilde{P}_3+2\tilde{P}_2+\tilde{P}_1)^3} \approx  \frac{1}{\langle W_{a|out}\rangle^2}\frac{(P_0\frac{27}{2} \sin^{4} 6gt+\frac{6}{16}\sin^2 6gt) \left(\frac{3}{4} \sin^{4} 3gt-\frac{3}{16} \sin^2 6 gt\right)^2 }{\left(P_0\frac{9}{4} \sin^{4} 6gt+\frac{3}{16}\sin^2 6gt+\frac{3}{2} \sin^{4} 3gt+\frac{3}{16} \sin^2 6 gt\right)^3}\nonumber\\
&\approx \frac{q_2(t)}{\langle W_{a|out}\rangle^2},\\
g^{(4)}_{a|out}(0)&\approx \frac{24 \tilde{P}_4}{(4\tilde{P}_4+3\tilde{P}_3+2\tilde{P}_2+\tilde{P}_1)^4} \approx  \frac{1}{\langle W_{a|out}\rangle^3} \frac{(P_0\frac{27}{2} \sin^{4} 6gt)\left(\frac{3}{4} \sin^{4} 3gt-\frac{3}{16} \sin^2 6 gt\right)^3}{\left(P_0\frac{9}{4} \sin^{4} 6gt+\frac{3}{16}\sin^2 6gt+\frac{3}{2} \sin^{4} 3gt+\frac{3}{16} \sin^2 6 gt\right)^4} \nonumber\\
&\approx \frac{q_3(t)}{\langle W_{a|out}\rangle^3},
\end{align}
and for $(6j+5)\pi/18 <gt<(6j+7)\pi/18$,
\begin{align}
g^{(2)}_{a|out}(0) &\approx \frac{1}{\langle W_{a|out}\rangle} \frac{(\frac{3}{2}\sin^4 3gt+\frac{6}{16}\sin^2 6 gt)(\frac{1}{16} \sin^2 6 gt-\frac{3}{4}  \sin^4 3gt)}{(\frac{3}{16}\sin^2 6gt+\frac{3}{2} \sin^4 3gt+\frac{3}{16} \sin^2 6 gt)^2}\approx \frac{q_1(t)}{\langle W_{a|out}\rangle},\\
g^{(3)}_{a|out}(0)&\approx  \frac{1}{\langle W_{a|out}\rangle^2}\frac{(P_0\frac{27}{2} \sin^{4} 6gt+\frac{6}{16}\sin^2 6gt) (\frac{1}{16} \sin^2 6 gt-\frac{3}{4}  \sin^4 3gt)^2 }{\left(P_0\frac{9}{4} \sin^{4} 6gt+\frac{3}{16}\sin^2 6gt+\frac{3}{2} \sin^{4} 3gt+\frac{3}{16} \sin^2 6 gt\right)^3}\approx \frac{q_2(t)}{\langle W_{a|out}\rangle^2},\\
g^{(4)}_{a|out}(0)& \approx  \frac{1}{\langle W_{a|out}\rangle^3} \frac{(P_0\frac{27}{2} \sin^{4} 6gt) (\frac{1}{16} \sin^2 6 gt-\frac{3}{4}  \sin^4 3gt)^3   }{\left(P_0\frac{9}{4} \sin^{4} 6gt+\frac{3}{16}\sin^2 6gt+\frac{3}{2} \sin^{4} 3gt+\frac{3}{16} \sin^2 6 gt\right)^4}\approx \frac{q_3(t)}{\langle W_{a|out}\rangle^3}.
\end{align}

\section{WC measurement by a mechanical oscillator}\label{OptomechanicsSI}
\subsection{Oscillator equations of motion}
The optomechanical coupling between the interferometer output field and a mechanical oscillator is described by the Hamiltonian \cite{Aspelmeyer2014RevModPhys}
\begin{align}
\hat H_{field+Osc}=\hbar\omega \hat n_{out}+\hbar \Omega \hat O^\dagger \hat O+G \hat n_{out}(\hat O^\dagger+\hat O),
\end{align}  
where $\hat O^\dagger (\hat O)$ are the creation (annihilation) operators of the mechanical oscillator, and $\hat n_{out}=\hat a_{out}^{\dagger} \hat a_{out}$ is the number operator for the output field. $\omega (\Omega)$ are the resonance frequencies of the output field (mechanical oscillator), and $G$ is the coupling strength between them. 

Defining the position and momentum operators for the oscillator to be
\begin{align}
\hat X=\frac{\hat O+\hat O^\dagger}{\sqrt{2}},\nonumber\\
\hat P=\frac{\hat O-\hat O^\dagger}{\sqrt{2}i}.
\end{align}
The Heisenberg equations of motion for these operators are
\begin{subequations}\label{EOMOsc}
\begin{align}
\frac{d \hat X }{d\tau}&=\Omega \hat P,\\
\frac{d \hat P }{d\tau}&=-\Omega \hat X-\sqrt{2} G \hat n_{out},\\
\frac{d \hat O^\dagger \hat O}{d\tau}&=-\sqrt{2}G \hat n_{out} \hat{P}
\end{align}
\end{subequations}
which have the solutions 
\begin{subequations}\label{SolutionOsc}
\begin{align}
\hat X (\tau)&=\frac{\sqrt{2}G (\cos\Omega \tau-1)}{\Omega}  \hat n_{out}+\hat X(0) \cos\Omega \tau+ \hat P(0) \sin\Omega \tau,\\
\hat P (\tau)&=-\frac{\sqrt{2}G\sin\Omega \tau}{\Omega} \hat n_{out}-\hat X(0) \sin\Omega \tau+ \hat P(0) \cos\Omega \tau,\\
\hat O^\dagger \hat O(\tau)&=\hat O^\dagger \hat O(0)+\frac{\sqrt{2}G}{\Omega}\hat n_{out}(\hat X(0)-\hat X(0)\cos\Omega \tau-\hat P(0)\sin\Omega \tau)-\frac{2G^2}{\Omega^2}\hat n_{out}^2(\cos\Omega \tau-1).
\end{align}
\end{subequations}
From the above equations, we calculate the variance in the displacement of the oscillator 
\begin{align}\label{variance}
\langle\Delta\hat X\rangle^2 (\tau)&= \langle\hat X^2(\tau)\rangle-\langle \hat X(\tau)\rangle^2,\nonumber\\
&=\left[\frac{\sqrt{2}G (\cos\Omega \tau-1)}{\Omega}\right]^2 (\langle (\hat n_{out})^2\rangle- \langle \hat n_{out}\rangle^2)+(\langle\hat X^2(0)\rangle-\langle \hat X(0)\rangle^2) \cos^2\Omega \tau \nonumber\\ 
&+(\langle\hat P^2(0)\rangle-\langle \hat P(0)\rangle^2) \sin^2\Omega \tau +(\langle \hat X(0) \hat P (0) \rangle+\langle \hat P(0) \hat X (0) \rangle-2\langle \hat X(0)\rangle \langle \hat P (0) \rangle) \cos\Omega \tau \sin\Omega \tau.
\end{align}

\noindent
\subsection{Oscillator in a coherent state:}
If the oscillator is initially in a coherent state $\ket{\alpha}$ with average phonon number $\bar{n}_O=|\alpha|^2$, we have 
\begin{align}
&\langle\hat X(0)\rangle=(\alpha+\alpha^*)/\sqrt{2},\langle\hat X^2(0)\rangle=(\alpha^2+\alpha^{*2}+1+2\bar{n}_O)/2,\langle\hat P(0)\rangle=(\alpha-\alpha^*)/\sqrt{2}i,\nonumber\\
& \langle\hat P^2(0)\rangle=-(\alpha^2+\alpha^{*2}-1-2\bar{n}_O)/2,\langle\hat X(0)P(0)\rangle+\langle\hat P(0)X(0)\rangle=(\alpha^2-\alpha^{*2})/i. 
\end{align}
For these values, Eq. \eqref{variance} becomes
\begin{align}
\langle\Delta\hat X\rangle^2 (\tau)=\frac{1}{2}+\frac{8G^2}{\Omega^2}\sin^4\left(\frac{\Omega\tau}{2}\right) (\langle (\hat n_{out})^2\rangle- \langle \hat n_{out}\rangle^2). 
\end{align}

For the output field emerging from the interferometer containing either cross-Kerr nonlinearity or even-$k$-photon exchange process, using Eqs. \eqref{VarErg} and (\ref{VarErgKeven}), we find
\begin{align}
\langle\Delta\hat X\rangle^2 (\tau)=\frac{1}{2}+ \frac{32G^2}{3\Omega^2}\sin^4\left(\frac{\Omega\tau}{2}\right) |\langle \Delta W_{a|out}\rangle^2|.
\end{align}
For small average photon number in the field, i.e., $\bar{n}_a\ll 1 $, using the relation $\langle \Delta W_{a|out}\rangle^2\approx 3\langle W_{a|out}\rangle-3\langle W_{a|out}\rangle^2$,  the above equation reduces to 
\begin{align}
\langle\Delta\hat X\rangle^2 (\tau)\approx \frac{1}{2}+ \frac{32G^2}{\Omega^2}\sin^4\left(\frac{\Omega\tau}{2}\right) (\langle W_{a|out}\rangle-\langle W_{a|out}\rangle^2).
\end{align}

From Eq. (\ref{SolutionOsc}c), the average phonon number evolves as
\begin{align}
\langle O^\dagger O\rangle(\tau)=\bar{n}_O+\frac{\sqrt{2}G}{\Omega}\langle\hat{n}_{out}\rangle\left(\frac{\alpha+\alpha^*}{\sqrt{2}}(1-\cos\Omega \tau)-\frac{\alpha-\alpha^*}{\sqrt{2}i}\sin\Omega \tau\right)+\langle (\hat n_{out})^2 \rangle \frac{4G^2}{\Omega^2}\sin^2\left(\frac{\Omega \tau}{2}\right).
\end{align}
For the output field from an interferometer containing either cross-Kerr or even-$k$-photon exchange,  using Eqs. \eqref{VarErg}  and \eqref{WCExp}, the above equation reduces to
\begin{align}
\langle O^\dagger O\rangle(\tau)&= \bar{n}_O+\frac{2\sqrt{2}G}{\Omega}\langle W_{a|out}\rangle\left(\frac{\alpha+\alpha^*}{\sqrt{2}}(1-\cos\Omega \tau)-\frac{\alpha-\alpha^*}{\sqrt{2}i}\sin\Omega \tau\right)\nonumber\\
&+\left[ \frac{|\langle \Delta W_{a|out}\rangle^2|}{3} +\langle W_{a|out}\rangle^2 \right] \frac{16G^2}{\Omega^2}\sin^2\left(\frac{\Omega \tau}{2}\right).
\end{align}

\noindent
\subsection{Oscillator in a thermal state:}
Let the oscillator be initially in a thermal state with average phonon number $\bar{n}_O$. Then we have 
\begin{align}
&\langle\hat X(0)\rangle=0,\langle\hat X^2(0)\rangle=(1+2\bar{n}_O)/2,\langle\hat P(0)\rangle=0,\nonumber\\
&\langle\hat P^2(0)\rangle=(1+2\bar{n}_O)/2,\langle\hat X(0)P(0)\rangle=i/2, \langle\hat P(0)X(0)\rangle=-i/2.
\end{align}
For these values, the position variance in Eq. \eqref{variance} becomes
\begin{align}
\langle\Delta\hat X\rangle^2 (\tau)=\frac{(1+2\bar{n}_O)}{2}+\left[\frac{\sqrt{2}G (\cos\Omega \tau-1)}{\Omega}\right]^2 (\langle (\hat n_{out})^2\rangle- \langle \hat n_{out}\rangle^2). 
\end{align}
For the output field emerging from the interferometer containing either cross-Kerr nonlinearity or even-$k$-photon exchange process, using Eqs. \eqref{VarErg} and (\ref{VarErgKeven}), we find
\begin{align}
\langle\Delta\hat X\rangle^2 (\tau)&=\frac{(1+2\bar{n}_O)}{2}+\frac{4}{3}\left[\frac{\sqrt{2}G (\cos\Omega \tau-1)}{\Omega}\right]^2 |\langle \Delta W_{a|out}\rangle^2|,\nonumber\\
&=\frac{(1+2\bar{n}_O)}{2}+|\langle \Delta W_{a|out}\rangle^2|\frac{32 G^2}{3\Omega^2}\sin^4\left(\frac{\Omega \tau}{2}\right).
\end{align}
For a small average photon number in the field, i.e., $\bar{n}_a\ll 1 $, we find $\langle \Delta W_{a|out}\rangle^2\approx 3\langle W_{a|out}\rangle-3\langle W_{a|out}\rangle^2$. Then the above equation reduces to 
\begin{align}
\langle\Delta\hat X\rangle^2 (\tau)\approx \frac{(1+2\bar{n}_O)}{2}+(\langle W_{a|out}\rangle-\langle W_{a|out}\rangle^2)\frac{32 G^2}{\Omega^2}\sin^4\left(\frac{\Omega \tau}{2}\right).
\end{align}

From Eq. (\ref{SolutionOsc}c), the average phonon number evolves as
\begin{align}
\langle O^\dagger O\rangle(t)=\bar{n}_O+\langle (\hat n_{out})^2 \rangle \frac{4G^2}{\Omega^2}\sin^2\left(\frac{\Omega \tau}{2}\right).
\end{align}

For the output field coming from the interferometer containing either cross-Kerr or even-$k$-photon exchange,  using Eqs. \eqref{VarErg}  and \eqref{WCExp}, we get
\begin{align}
\langle \hat O^\dagger \hat O\rangle(\tau)= \bar{n}_O+\left[ \frac{|\langle \Delta W_{a|out}\rangle^2|}{3} +\langle W_{a|out}\rangle^2 \right] \frac{16G^2}{\Omega^2}\sin^2\left(\frac{\Omega \tau}{2}\right).
\end{align}
For $\bar{n}_a\ll 1$, using the relation $\langle \Delta W_{a|out}\rangle^2\approx 3\langle W_{a|out}\rangle-3\langle W_{a|out}\rangle^2$, the above equation reduces to
\begin{align}
\langle \hat O^\dagger \hat O \rangle(\tau)\approx \bar{n}_O+ \langle W_{a|out}\rangle \frac{16 G^2}{\Omega^2}\sin^2\left(\frac{\Omega \tau}{2}\right).
\end{align}

\vspace{2cm}

\providecommand{\newblock}{}

\end{document}